\documentclass[fleqn,usenatbib]{mnras}
\usepackage{enumitem}

\usepackage[T1]{fontenc}

\DeclareRobustCommand{\VAN}[3]{#2}
\let\VANthebibliography\thebibliography
\def\thebibliography{\DeclareRobustCommand{\VAN}[3]{##3}\VANthebibliography}

\usepackage{graphicx}	
\usepackage{amsmath}	

\usepackage{float}
\restylefloat{table}
\usepackage{txfonts}
\usepackage{xcolor}
\usepackage{subcaption}
\usepackage{color, colortbl}
\definecolor{Gray}{gray}{0.9}
\definecolor{LightCyan}{rgb}{0.88,1,1}
\usepackage[para,online,flushleft]{threeparttable}
\title[Machine learning classifiers for AGN]{Automated algorithms to build Active Galactic Nuclei classifiers}
\author[S. Falocco et al.]{
  S. Falocco,$^{1,3}$\thanks{E-mail: falocco@kth.se, $\rm{agn_{-}ml}$@virgilio.it }
   F.J. Carrera$^{2}$
and J. Larsson$^{1}$
\\
$^{1}$KTH Royal Institute of Technology, AlbaNova, SE-106-91 Stockholm, Sweden\\
$^{2}$Instituto de F\'\i{}sica de Cantabria (CSIC-UC), Avenida de los Castros, 39005 Santander, Spain\\
$^{3}$Nexer Insight AB, Regeringsgatan 29, 11153 Stockholm, Sweden
}

\date{Accepted XXX. Received YYY; in original form ZZZ}

\pubyear{2021}
\begin{document}
\label{firstpage}
\pagerange{\pageref{firstpage}--\pageref{lastpage}}
\maketitle

\begin{abstract}
   We present a machine learning model to classify Active Galactic Nuclei (AGN) and galaxies (AGN-galaxy classifier) and a model to identify type 1 (optically unabsorbed) and type 2 (optically absorbed) AGN (type 1/2 classifier). 
  We test tree-based algorithms, using training samples built from the X-ray Multi-Mirror Mission -Newton (\emph{XMM-Newton}) catalogue and the Sloan Digital Sky Survey (SDSS), with labels derived from the SDSS survey.
 The performance was tested making use of simulations and of cross-validation techniques.
    With a set of features including spectroscopic redshifts and X-ray parameters connected to source properties (e.g. fluxes and extension), as well as features related to X-ray instrumental conditions, the precision and recall  for AGN identification are 94 and 93 per cent, while the type 1/2 classifier has a precision of 74 per cent and a recall of 80 per cent for type 2 AGN. The performance obtained with photometric redshifts is very similar to that achieved with spectroscopic redshifts in both test cases, while there is a decrease in performance when excluding redshifts.
    Our machine learning model trained on X-ray features can accurately identify AGN in extragalactic surveys. The type 1/2 classifier has a valuable performance for type 2 AGN, but its ability to generalise 
    without
    redshifts is hampered by the limited census of absorbed AGN at high redshift.
\end{abstract}




\section{Introduction}

   Active Galactic Nuclei (AGN) are the brightest persistent sources in the sky and their luminosities emitted in their innermost compact regions are a sign of accretion onto supermassive black holes.
   X-ray surveys have helped to expand our knowledge of AGN, challenging the sensitivity limits and continuously expanding the AGN census  (e.g., \citealt{nandra2014,luo2016,hasinger2021}). The advantage of X-ray surveys is that they allow us to detect sources obscured in other wavelengths, since X-ray spectra emitted from the nuclei do not strongly interact with surrounding material. Telescopes such as X-ray Multi-Mirror Mission - Newton (\emph{XMM-Newton}; \citealt{xmm}) have helped to identify  a larger number of AGN, including absorbed and low-luminosity AGN, helping to distinguish them from galaxies without an active nucleus (e.g., \citealt{caccianiga2007}, \citealt{elias2021}, \citealt{torbaniuk2021}). On the other hand, optical surveys such as the Sloane Digital Sky Survey SDSS \citep{york2000} have provided  information about identifications and redshifts over $\sim30$ per cent of the sky. The survey provides, besides multi-colour data, also spectroscopic information which gives information on absorption along the line of sight, with broad optical lines observed in unabsorbed AGN (also called type 1 AGN) and not typically detected in absorbed AGN (also called type 2 AGN), the difference being commonly interpreted as an orientation effect of the absorbing material.
   
   Large surveys are planned for the next years and the incoming huge data sets will bring information about galactic and extragalactic sources. For instance, in the field of optical - near infrared astronomy, the Legacy Survey of Space and Time (\emph{LSST}, \citealt{lsst2019}) will bring a dramatic improvement in AGN demography  across cosmic times: with its wide and deep field observations, it will provide data of millions of sources per night. The data flows will be beyond the limits of human ability to analyse and the volume of the archives will grow fast. Classifying sources in these conditions is a challenging task and, given the complexity of the algorithms, they cannot be built manually. Automated tools become useful to analyse such large data sets, but also in a more general context of moderately-sized surveys, since they have the power to quickly perform the desired tasks (e.g., classification or prediction) for new data flows.  
 The expression machine Learning (ML) refers to building automated procedures which directly learn from data in order to perform the above-mentioned tasks \citep{samuel}.
 A specific branch of ML is supervised learning, which makes use of labeled data, i.e. data with the desired outputs. These techniques provide the machine with instructions for how to process the information behind labeled data in order to learn a model from them. Finding a relationship between a set of measurements (features) and a target variable (label), ML methods use this relationship or model to predict the target variable for new data.

 Supervised learning algorithms are broadly used in different fields where the labels are commonly referred to as 'ground truth', since they are trusted identifications, while features usually have some degree of uncertainty \citep{mitchell97}. The techniques developed so far are suitable whenever the level of uncertainty and noise are similar across the features \citep{Zhu2004}. Techniques of data cleaning are usually employed to filter noisy samples (e.g., \citealt{noisefilter}, \citealt{sanchez2003}), while dimensionality reduction can help reduce the number of features by ignoring those that bring redundant information or noise
 \citep{pearson, tipping1999}.
 
 ML has successfully been used in astronomy in several contexts, for instance to identify transient sources \citep{disanto2016} and AGN using their optical variability \citep{decicco2021} or their magnitudes in optical surveys  \citep{cavuoti}. ML methods have also allowed to address other aspects of extragalactic surveys, such as to estimate redshifts: for instance,  \cite{mountrichas2017} estimated photometric redshifts of more than 1000 X-ray sources. Redshifts were also estimated in \cite{ruiz2018} for more than 50 per cent of the 3XMM catalogue, with an outliers rate ranging from 4 to 40 per cent, depending on the amount of available data. For the classification of X-ray sources, ML methods have been applied to classify sources in the 4XMM-DR9 catalogue by \cite{zhang2021}, additionally using infrared data. The authors classified galaxies, quasars and stars with a total accuracy of around 95 per cent.  ML has also been valuable for expanding the census of quasars at high redshift \citep{lukas2021}.
   Finally, the high predictive power of ML methods enables assessment studies for new instruments, for instance forecasting background contamination for the future X-ray mission \textsl{Athena} \citep{kronberg2020}.
 
The aim of this work is to determine the performance of ML methods to identify sources in X-ray surveys, investigating the relevance of features from X-ray and optical surveys. In view of data from recent facilities such as \textsl{eRosita} \citep{merloni2012} and future X-ray observatories such as \textsl{Athena} \citep{nandra2013}, this work aims at exploring frameworks and building models that can be deployed for incoming data.
The aim of this paper is twofold: in the first part we build a supervised learning algorithm to identify AGN and galaxies (test case 1); a second algorithm aims at distinguishing between two subclasses of AGN, specifically unabsorbed (type 1) AGN and absorbed (type 2) AGN (test case 2). The algorithms are built from training on two of the largest existing data sets in the archives, namely the \emph{XMM-Newton} and SDSS surveys (details about the data are provided in Section \ref{data}).
For our classification tasks, the features are derived from the \emph{XMM-Newton} X-ray catalogue and the labels from SDSS (in test case 1) and the Million quasar survey 
(combined with SDSS and XMM-Newton,
in test case 2).

   This paper is organised as follows: data and sample construction are described in Section \ref{data}; methods are described in Section \ref{methods}; results are discussed in Section \ref{discussion}, conclusions and future applications are described in Section \ref{conclusions}. Additional information on the samples used to train the models are provided in Appendix \ref{appdatadetails}, while results on the AGN-galaxy classifier trained after having relabeled high-luminosity galaxies are described in Appendix \ref{relabelledAGNgalaxies}, a test to check the impact of features is shown in Appendix \ref{physfeatimportance}.

   \section{Data collection}\label{data}
   
The first catalogue involved in this study is the \emph{XMM-Newton} Catalogue. We used the 4XMM DR9 19/12/2019 detection catalogue presented in \cite{webb2020}\footnote{See \url{http://xmm-catalog.irap.omp.eu}}, referred to as 'XMM' hereafter.
The ML methods used in this paper are based on supervised learning, which makes use of labeled data from existing catalogues in order to build a classifier. 
For the galaxy - AGN identification, which is the first step of this work,  the classifications (i.e., the labels) have been derived from optical spectra in SDSS data release 16 (2019, \citealt{aguado2019}). Redshifts and labels are therefore obtained from the DR16 SDSS spectral information\footnote{ \url{https://www.sdss.org/dr16/spectro/spectro_access/}}, which results in three classes: AGN, galaxies and stars.

The type 1 - type 2 AGN  identifications are derived from the Million quasar survey: Version 7.0, 30 September 2020 (Milliquas)\footnote{\url{https://heasarc.gsfc.nasa.gov/W3Browse/all/milliquas.html}} in \cite{flesch2019}.
This catalogue mainly takes information from SDSS, and in part also from  AllWISE \citep{secrest2015}. 
The procedure used to build the training sample is illustrated in Fig. \ref{sampleconstruction} and described below. More details on sample construction are included in Appendix A and Fig. \ref{samplecomposition}.

\subsection{AGN and galaxies from XMM and SDSS}
\label{xmm-sdss}

   \begin{figure}
     \includegraphics[width=0.52\textwidth, height=15cm]{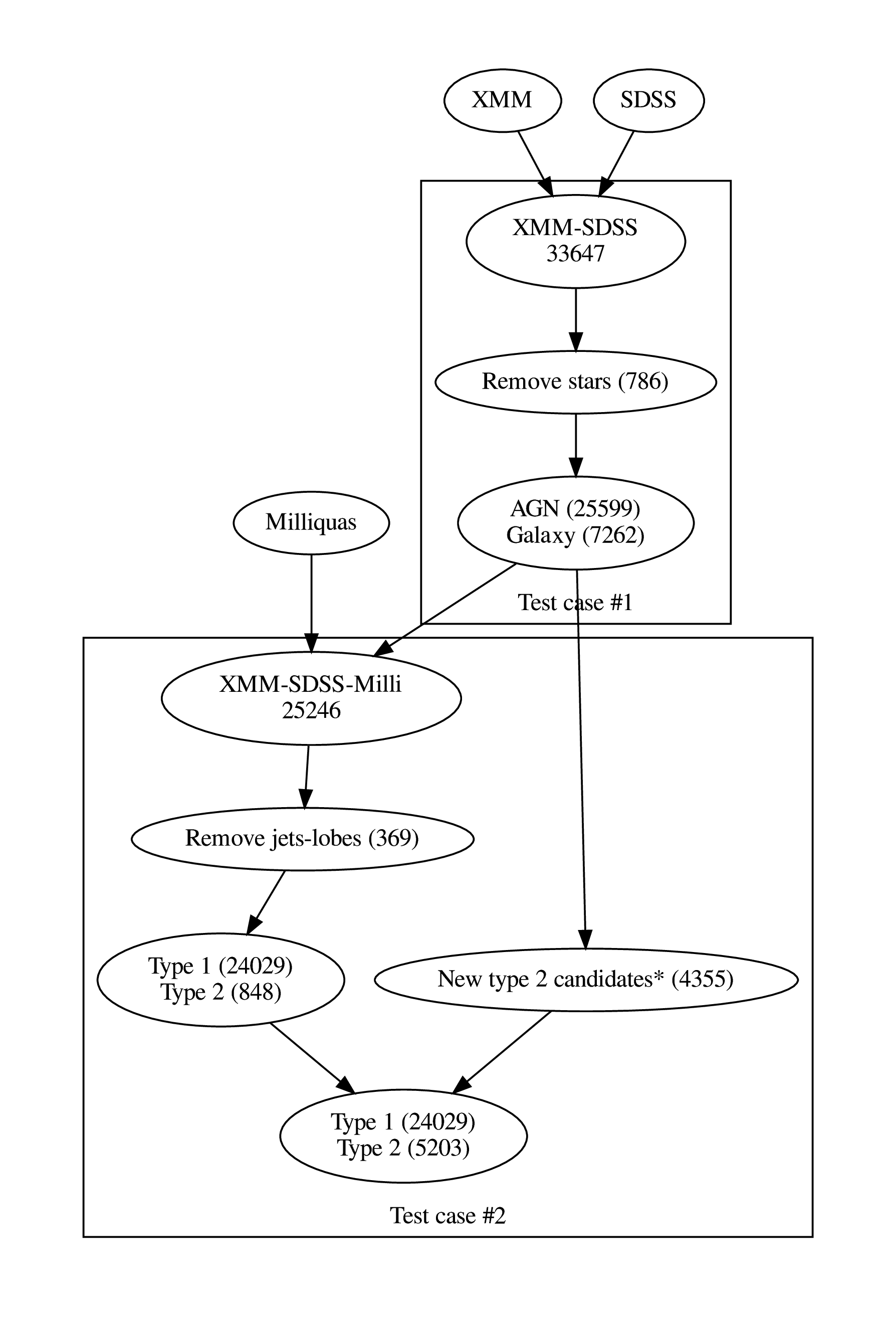}
       \caption{Sample construction in the two test cases. The two squares enclose, respectively, test case 1 to identify AGN and galaxies, and test case 2 to divide the AGN class into type 1 and type 2 AGN. The sample construction of XMM-SDSS is described in Section \ref{xmm-sdss}, while the construction of XMM-SDSS-Milli is described in Section \ref{xmm-sdss-milli}. New type 2 candidates*:  galaxies not included in Milliquas, with $L>10^{42}~ \rm erg~s^{-1}$ (see text for details). More details of the sample construction are illustrated in Fig. \ref{samplecomposition}.}
         \label{sampleconstruction}
   \end{figure}

 As a first step to construct our sample, a catalogue of sources in common between XMM and SDSS was constructed (using the full table SpecObj of SDSS). This was done matching the coordinates of XMM and SDSS at 5 arcsec.  We selected the closest match (in sky distance) if more than one SDSS source was within the 5 arcsec radius. The cross-correlation radius is the same used in many previous works such as \cite{falocco2014}, \cite{traulsen2019} and \cite{webb2020}; its value is limited by the angular resolution of X-ray instruments: the PSF has Full Width Half Maximum of 6 arcsec and Half Energy Width of 15 arcsec\footnote{See Section 3.2.1.1 of \emph{XMM-Newton} Users Handbook} in \emph{XMM-Newton}.
   In this step we obtained 33647 detections (XMM-SDSS sample hereafter), corresponding to 33469 individual sources. The starting X-ray catalogue contains all detections in XMM, 810795 in total, that have not been filtered.
 Though in the XMM-SDSS sample the difference between the number of detections and the number of sources is only $\sim$200, it is useful to exploit all available detections for each source. This is because each of them results in different observational features due to variability of the source itself as well as different observational conditions. Including a variety of observations of the same source in the training sample helps to train a classifier on a broad range of conditions that would not be captured if only one observation for each source was selected.

 AGN objects are defined extracting labels from SDSS using the columns 'class' and 'subclass' in that catalogue. In SDSS the class can assume three values; QSO, galaxy and stars. The sources labeled as stars have been excluded from this work because we focus on exploring ML classification of extragalactic surveys. With respect to this, we note that a simple cut in latitude (e.g. between -30 and 30 degrees) would remove 72 per cent of stars if this label had been unavailable. 
The AGN class is composed of sources with evidence for a compact nucleus as per the SDSS catalogue\footnote{AGN were selected based on two conditions: 1): CLASS = QSO; 2): CLASS = GALAXY and (SUBCLASS = AGN or SUBCLASS = AGN BROADLINE)};
the class of galaxies is also based on the SDSS classification\footnote{Galaxies were selected based on the condition: CLASS = GALAXY and SUBCLASS other than AGN and AGN BROADLINE (they might have no subclass provided in SDSS or STARFORMING, or STARBURST)}.
The resulting catalogue is composed of detections divided as follows:
\begin{itemize}
 \item 25599 AGN, 
\item 7262  galaxies
  \end{itemize}
  
The definition of galaxies may in principle have some mislabeling due to the fact that very faint AGN could be difficult to identify from the optical spectra. This could for instance happen for obscured nuclei or intrinsically faint AGN, also called Low Luminosity AGN (LLAGN). In order to better investigate this aspect, we
crossmatched the full sample of AGN and galaxies with Milliquas, which contains extra information in addition to that included in SDSS (AllWISE is also used as mentioned above). Details of the crossmatch with Milliquas are reported in Section \ref{xmm-sdss-milli}.

Finally, we created a version of the AGN-galaxy sample with photometric redshifts, which allows us to test how the type of redshift information affects the ML classification.
The photometric redshifts were obtained from the Kilo-Degree-Survey (KiDS) by \cite{kuijken2019}\footnote{\url{http://kids.strw.leidenuniv.nl/DR4/}}. Crossmatching the sample of 25599 AGN plus 7262 galaxies with $\sim$ 2 million KiDS sources, which includes quasars from \cite{nakoneczny2021} and bright galaxies from \cite{bilicki2021}, we obtained 260 galaxies and 710 AGN with photometric redshifts. The crossmatch was done between the optical SDSS coordinates and the KiDS coordinates using a 1 arcsec radius. This value is widely used in the literature as a reasonable matching radius for optical catalogues (e.g., \citealt{falocco2015} and \citealt{decicco2019}). The photometric sample constructed in this way has a similar redshift distribution as the initial spectroscopic sample, similar selection effects in redshift, as well as a similar degree of imbalance between the majority and minority classes.
The results of training the ML algorithm on this sample and using photometric instead of spectroscopic redshifts are reported in Section \ref{discussionAGNgalaxies}.

\subsection{AGN from XMM, SDSS and Milliquas}
\label{xmm-sdss-milli}
We further investigate the AGN-galaxy sample defined above by cross-correlating it with Milliquas.
Matching the coordinates at at a radius of 1 arcsec (value motivated above) results in 25246 extragalactic sources in common between the parent surveys (see Fig. \ref{samplecomposition} in Appendix~A and Fig.~\ref{sampleconstruction}). 
Out of these sources, 24691 were defined as AGN from the SDSS survey and  555 were defined as galaxies in SDSS (as explained in the previous section). These  555 SDSS galaxies clearly show some evidence for being AGN since they are in the Milliquas sample. They represent $\sim7$ per cent (555 out of 7262) of sources identified as galaxies in SDSS. The effect of relabeling these 555 sources on the classification performance will be described in Section \ref{discussionAGNgalaxies}. An opposite problem could arise if sources identified as AGN in SDSS have no nucleus, but this is expected to have a lower incidence, given that optical spectroscopy is one of the most solid methods to find and classify AGN.

There is a group of $\sim$900 sources labeled as AGN in SDSS which do not appear in Milliquas, most probably because the Milliquas survey removed all sources that have a probability of being AGN lower than 80 per cent, thus excluding faint SDSS objects with low S/N spectra. This resulted in removing over 20k SDSS objects in the Milliquas survey, as explained in Section 2 of \cite{flesch2019}. We call the full sample of 25246 extragalactic sources XMM-SDSS-Milli.

\subsection{Type 1  - Type 2 AGN from XMM-SDSS-Milli}
\label{type1type2trainingsample}
Among XMM-SDSS-Milli, we further selected only those sources identified as type 1 or type 2 following the identifications in the Milliquas catalogue. Other kinds of sources like lobed or jetted AGN (369 sources) have not been included in this sample to explore the potential of ML classification in finding type 2 AGN versus type 1 AGN. 
Type 1 AGN are those with evidence for an unabsorbed nucleus in the optical and Type 2 include those with evidence for an obscured nucleus\footnote{Type 1 are defined as 'q' (QSO, type-I broad-line core-dominated) and 'a' (AGN, type-I Seyferts/host-dominated) in the Milliquas survey, while
Type 2 AGN have labels 'K' (NLQSO, type-II narrow-line core-dominated) and 'N' (LAGN, type-II Seyferts/host-dominated).}.
This selection results in 24877 observations of:
\begin{itemize}
\item 24029 type 1
\item 848 type 2
\end{itemize}

We added to this sample new type 2 candidates from the parent XMM-SDSS sample. These candidates were not included in the parent Milliquas survey and labeled as galaxies in SDSS; their X-ray luminosities\footnote{The luminosity is determined as $F \times 4~\pi~d^{2}$ where $F$ is the flux and $d$ is the luminosity distance extracted from the redshift using the cosmological parameter $H_{0}=72~\rm km~s^{-1} Mpc^{-1}$.} are above  $10^{42}\ \rm  erg~s^{-1}$ (between 0.2 and and 12~keV, corresponding to band 8 of the XMM catalogue).
Literature reports convincing evidence that galaxies with X-ray luminosities higher than this threshold indeed have some kind of AGN activity \citep{ranalli2003,maiolino2003,castello-mor2012}.
There are 4355 optical galaxies excluded from the Milliquas survey with X-ray luminosities above the limit of $10^{42}\ \rm  erg~s^{-1}$ 
that can be interpreted as extra candidate type 2 AGN. These sources are classified as galaxies in the optical band, but most likely have obscured nuclei as attested by their high X-ray luminosities.
Test case 2 is therefore based on a sample of 29232 observations of:
\begin{itemize}
\item 24029 type 1
\item 5203 type 2
\end{itemize}

This sample is used for training the type 1 type 2 AGN (type 1/2) classifier. The sample and its composition is detailed in Fig. \ref{samplecomposition}. It includes 24476 sources defined as AGN in SDSS and 4756 sources defined as galaxies in SDSS.

In test case 1 we have 7262 SDSS sources classified as galaxies in SDSS (as detailed in the Section \ref{xmm-sdss-milli}), but 555 of them have evidence (from Milliquas) for an AGN nucleus.
On the other hand, there are 4355 SDSS galaxies that do not have evidence for a nucleus from optical spectroscopy, but where active nuclei can be inferred from their high X-ray luminosities. We found that these sources have a small impact on the final results on the AGN-galaxy classification, as will be explained in Section \ref{discussionAGNgalaxies} and Appendix \ref{relabelledAGNgalaxies}. 

As for test case 1, we performed an additional experiment also for the type 1/2 classifier in order to explore its performance when substituting spectroscopic with photometric redshifts. This was done crosscorrelating the optical coordinates of the above sample of type 1 and type 2 AGN with the KiDS survey as described in Section \ref{xmm-sdss} for test case 1. The starting sample of 24029
type 1 AGN plus 5203 type 2 AGN crossmatched with the KiDS sources results in 687 type 1 AGN and 155 type 2 AGN with photometric redshifts. This photometric sample has a very similar imbalance and selection effects in redshifts as the corresponding spectroscopic sample, as we also noted in test case 1. The results of training the ML algorithm on this sample using photometric redshifts are reported in Section \ref{resultsAGNtype}.


\section{Methods}\label{methods}
The workflow follows the main steps detailed in this section:
\begin{enumerate}[label={\arabic*.}]
\item Features selection (see Section 3.1 below)
\item Define the ML algorithms most suitable for the data (described in Section 3.2)
\item Estimate the performance of the classifiers with the 10 fold cross validation (see Section 3.3)
  \item Features importance (see Section 3.4)
\end{enumerate}
   This work used the following \small{PYTHON} \normalsize libraries:  \small{SKLEARN} \normalsize for the ML algorithms application and feature importance \citep{scikit-learn}; \small{IMBALANCED-LEARN} \normalsize \citep{imblearn} was used to apply oversampling techniques as described below. 
 Catalogues have been handled with \small{PANDAS} \normalsize \citep{mckinney}; most plots have been made with \small{MATPLOTLIB} \normalsize \citep{matplotlib}, some plots made use of \small{SEABORN}\normalsize, presented in \cite{seaborn}.
  Decision tree-based algorithms make use of probabilistic decisions and some degree of randomness (i.e. when a feature is selected to split the sample in a tree-based model). For this reason, every time a ML algorithm is trained and evaluated over the same data set, it will give slightly different results.
  In order to reproduce exactly the same numbers at every run, we fixed the random seed in \small{PYTHON} \normalsize to 42. We note that changing the random seed would cause a change in the performance metrics of less than 0.1 per cent, and all results described below are unaffected by this.

   \subsection{Features selection}\label{featureselection}

   Keeping the dimension of the training set small is useful when applying ML algorithms with the purpose to both reduce the computing time and to avoid feeding the ML algorithm with redundant features.
   The features used to train the classifiers (defined in Table \ref{tab:featuresdescription}) are selected with the procedure described below.

 The training samples have columns from the XMM catalogue, besides the optional redshifts and the labels.  We defined the features set after a first filtering of the columns.  
 First of all, we only used columns whose values were present for all sources in the sample,
 which left 94 columns. We preferred to filter columns, not rows, in order to have a reasonable size of the dataset (filtering rows with missing values would leave only a few sources in the sample).
    Strongly   correlated  features 
   (defined as having a Pearson correlation coefficient with a p-value P$\geq0.9$) have been eliminated, because they give redundant information.

A group of features related to source properties are retained in our selection.
Firstly, we decided to keep the flux in the full X-ray band, namely band 8 in the catalogue (0.2 -  12.0 keV), for the AGN-galaxy classification, because it can provide hints for an active nucleus. These full-band fluxes are often used to check for the presence of an AGN in a galaxy \citep{ranalli2003,castello-mor2012}.
   This was not done for the type 1/2 classification because the individual bands can be more instructive when searching for type 1 and type 2 AGN. Specifically, the  fluxes
   below 2 keV and  above 2 keV   have proven useful to select type 2 AGN  (e.g., \citealt{terashima2015}) and to train ML algorithms \citep{farrell2015}.  
  Fluxes in bands  2, 3 and 4  of XMM have been combined into a single value covering the energy range between 0.5 and 4.5 keV.
 The total number of counts in the band between 0.2 and 12 keV  ($EP_{-}8_{-}CTS$) has been included in both test cases, as well as the net exposure time of the observation ($EP_{-}ONTIME$). These two parameters are relevant because they give information on how many X-ray counts can be collected in a given exposure time (i.e. their ratio is the source count-rate). 
We further included $SC_{-}DET_{-}ML$ in the full band because sources with different redshifts may have different detection likelihoods (e.g., at high redshifts, there is a higher probability to detect high-luminosity sources rather than faint sources).
  We also include parameters related to the extent in the merged observation; $SC_{-}EXTENT$  and $SC_{-}EXT_{-}ML$. We prefer to use these features from the merged observation rather than the individual ones because they are more robust (the extent is not expected to change across the observations).

  Features related to instrumental conditions are included in a first run and then excluded in a second run to quantify their actual influence on the the trained model.
 $SUM_{-}FLAG$ and $N_{-}DETECTIONS$ are part of this group: while the first one is a summary of the quality flags in each camera, the second one gives the number of detections of each unique source, and thus how many detections contribute to the merged parameters.
The distance from the nearest neighbours ($DIST_{-}NN$) is also used in order to check its relevance in the classification, since it has been included in previous works \citep{farrell2015}; the offaxis angle ($EP_{-}OFFAX$) has also been kept to check if it might affect the classification (in principle it might affect the extension, thus the ability to distinguish genuine extent from artefacts). We will see, however, that $EP_{-}OFFAX$ has no influence in our ML algorithms, so our classification tasks are not affected by this parameter.
We used spectroscopic redshifts in a first run; photometric redshifts have been used to replace them in a second run; the last training run was done without redshifts.

The selection just described includes the features listed in Table \ref{tab:featuresdescription}. As can be noted, there are directly source-related properties (from feature 1 to 8 of the table) and other features more connected with the instrumental conditions (i.e., from feature 9 to the end of the list in the same table). We demonstrate the consistency of results by retraining our models including all of the features, and only the directly source-related properties, as described in the \ref{physfeatimportance}.
Our feature selection just described results in 13 columns for the AGN-galaxy classification (those defined in Table \ref{tab:featuresdescription}) and 12 columns in the type 1/2 classification. The reason for this difference is that we decided to add the flux in the full band, $EP_{-}8_{-}FLUX$, in test case 1, following the literature as mentioned above.

\begin{table*}[h]
  \begin{threeparttable}
 \begin{tabular}{l|l|l}
   \hline
      &   Parameter   & Definition  \\

1 	&	 $SC_{-}EXTENT$ 	&  Extent in the merged observations  in arcsec \\ 
2 	&	 $SC_{-}EXT_{-}ML$ 	& Average of extent likelihood in the detections of the source\tnote{1}\\ 
3 	&	 $EP_{-}8_{-}FLUX$ 	&  Flux in EPIC in 0.2-12 keV \tnote{2}\\ 
4 	&	 $EP_{-}FLUX_{-}2_{-}3_{-}4$ 	& Sum of EPIC fluxes  in 0.5-1, 1-2, 2-4.5 keV	\tnote{2}\\ 
5 	&	 $SC_{-}DET_{-}ML$ 	&  Maximum detection likelihood in band 0.2 - 12 keV\\ 
6 	&	 $EP_{-}5_{-}FLUX$ 	&   EPIC flux in band 4.5-12 keV \tnote{3}	 \\ 
7 	&	 $EP_{-}8_{-}CTS$ 	&  EPIC combined counts in band 0.2-12 keV \\ 
\hline
8 	&	 $Z$ 	&  Spectroscopic redshifts	  \\ 
\hline
9 	&	 $DIST_{-}NN$ 	&    Distance to the nearest neighbour detection (arcsec)\\
10 	&	 $EP_{-}ONTIME$ 	&  Net exposure  time after filtering (s)\\ 
11	&	 $SUM_{-}FLAG$ 	&  Quality flag\tnote{3}\\ 
12	&	 $EP_{-}OFFAX$ 	& Offaxis angle (arcsec)\\ 
13	&	 $N_{-}DETECTIONS$ 	&  Number of detections in merged observations \\ 
 \end{tabular}
 \begin{tablenotes}
  \item[1] The extent likelihood is defined in XMM as $-\ln{p}$, where p is the probability of the extent occurring by chance; \item[2]  Flux units: $\rm erg~cm^{-2}~s^{-1}$;
   \item[3] Sum of the quality flags of individual cameras, it can assume zero value for the most reliable detections and values increasing until 4 in the worse cases when the source is most likely spurious.
   \end{tablenotes}
\caption{Features used to train the classifiers. All features are used for the AGN-galaxy classification, while $EP_{-}8_{-}FLUX$ is not included in the type 1/2 classifier training. All features except for redshifts are from the XMM catalogue.}
 \label{tab:featuresdescription}
  \end{threeparttable}
  \end{table*}

   \subsection{Supervised learning algorithms}\label{algorithms}

   Several models have been used in this work for classification. The subset of the survey used to fit the classifier is referred to as training sample, while the subset used to test the sample is composed by different datapoints not included in the training sample, which constitute to the so-called test sample.
  
   The methods tested in this work are tree-based algorithms, which are a sequence of if-else conditions to split the survey. This process is optimised in order to build meaningful splits. This is done minimising the Gini impurity \citep{breiman1984}. 
 The Gini impurity for a split is the probability for a datapoint to be identified with the right label $i$ $p(i)$ multiplied for the probability to be misclassified ($1-p(i)$).
The formula of the Gini impurity is:
\begin{equation*}
 \sum_{i=1}^{i=c} p(i)(1-p(i))
  \end{equation*}
which sums over the classes in the data set.
An ideal perfect split would give a Gini impurity of 0 and can be achieved if all datapoints in the split fall into a single class. We point out that the Gini criterion is not used to evaluate the goodness of a ML model. Instead, it is used only to estimate how homogeneous a split is. The Gini criterion is used to optimise the sample splitting in the decision tree: it determines the selection of features and threshold to define the nodes and the leaves of the tree, as will be explained below.

   In ensemble methods such as random forest \citep{breiman2001}, the decision tree represents only a starting point to develop better algorithms.  This is due to the fact that the simple decision tree could result in overfitting. Overfitting is detected when the  performance of the training set is excellent while the performance of the test dataset is poor, and this occurs because the model has learnt noise patterns in the training dataset. We have taken several measures to avoid overfitting: keeping the dataset dimension low,  N fold cross validation to properly estimate performance metrics, and making use of ensemble algorithms.
   The 10 fold cross validation \citep{stone1974} is described in the next section and the features selection has been explained in the previous section, below we describe the algorithms that we have used and which help to address the problem of overfitting.
   
   An advantage of tree based algorithms is that they do not make specific assumptions on the data distribution, as for instance the mean and the standard deviation for a normal gaussian distribution \citep{friedman1977}.
   On the contrary, these models learn from the data with high flexibility. For this reason we do not use the Naive Bayes algorithm, that instead assumes a gaussian distribution of the data in the parameter space (e.g. \citealt{zhang04}). We did not explore in detail Support Vector Machine approaches either \citep{crammer2001}, because they might not be suitable for data sets where the classes overlap in the parameter space. This is expected in our classification problem, since there is a continuous distribution between AGN and galaxies; indeed, LLAGN and Low Ionisation Nuclear Emission Line Regions have intermediate luminosity between AGN and galaxies. 
   There is also a continuous distribution between type 1/2 AGN  \citep{hasinger2008}.
   
   Below, we briefly describe the ML algorithms used in this work.
   \begin{itemize}
   \item [Tree] The decision tree is a sequence of if-else conditions to split a labeled sample. Optimisation methods are used to build meaningful splits, estimating the homogeneity of the labels in the subsets. The splitting is continued until a stopping criterion is met, which prevents overfitting. In our setting of hyperparameters, the optimisation criterion is the Gini impurity. The stopping criterion adopted is to have a minimum of 100 sources in each split or a 
   maximum
   tree depth of 10. The tree depth is the number of iterations of the decision tree. The decision tree algorithm was presented in \cite{breiman1984}
\item [RF] Random Forest starts from a set of decision trees to build a better classifier starting from the idea that a large number of uncorrelated trees will outperform any of the individual tree models. The reason for this behaviour is that each tree will make errors in a different direction and some trees will perform better than others, as a consequence their combination might improve the result. 
  Random Forest builds a forest of trees trained on different random sets, drawn from the training samples with the bootstrap method. Each bootstrap sample given as input to the decision tree is not a chunk of the training dataset, but it 
  instead
  has the same size as the training sample. A bootstrap sample is composed by a number U of unique datapoints from the initial training sample  of dimension $N$, the rest $N-U$ of the bootstrap sample are duplicates. 
 Moreover, each tree uses a random subset of the features, composed by $f$ features. This is a parameter of the model, for instance it can be equal to the total number of features $n$, $\log_{2}(n)$, $\sqrt{n}$, or even an arbitrary number. 
 Randomness introduced by the choice of features and the choice of the bootstrap samples is a benefit of this method, since it is able to produce different individual trees. The combination of these individual, randomly chosen trees, is less prone to overfitting than one individual deep tree.
 In our specific settings, the number of features $f$ is $\sqrt{n}$ and the number of weak learners (decision trees) in the forest has been set to 100.  Random forests were presented in \cite{ho1998} and \cite{breiman2001}
\item [AB] The AdaBoost algorithm starts from a set of simple decision trees and acts iteratively to improve the resulting classification. The algorithm is designed to improve the performance at each iteration.
 The algorithm is built such that, in each subsequent iteration, the incorrectly identified data will have an increased weight, while the correctly identified ones will have a lower weight so that the model will focus more on the most difficult data. The final prediction is therefore the weighted average (or the mode of the labels in classification problems) of the predictions obtained from the previous tree models. In this work, it is built with 100 weak learners (simple decision trees with depth = 1). This algorithm was presented in \cite{freund1997}
\item [grad] Gradient Boosting is another iterative algorithm that trains weak learners in sequence, similarly to AB.  The main difference between the two algorithms is that grad does not use the weights to train the new model at each iteration. It uses instead the optimisation of a loss function, a function which measures the error rate in the classification. In our settings, the weak learners are again 100 simple decision trees (with depth=1). The learning rate (an optimisation parameter which indicates the gain of the model) has been set to 1.0. This algorithm was presented in \cite{friedman2001}
\item [vote]  Voting algorithm with different weights (w) attributed to the estimators. This algorithm takes the weighted majority vote among these estimators. We initially setup the algorithm such that AB (w=2), Tree (w=1), RF (w=1), grad (w=2).
  The weights were chosen in this way to consider preferentially the boosting algorithms AB and grad which are most often free from overfitting.
However, our ML algorithms have a similar performance and converge to the best performing tree-based algorithm, as will be discussed in Section \ref{discussion}, thus the specific weights of each of them are not relevant. We have indeed noticed that adopting the same weight = 1 for all estimators does not affect the results. Therefore, to simplify the discussion, we report the results with equal weights for all estimators in the next sections.
  The voting algorithm was presented in \cite{zhang2014}
   \end{itemize}
Different hyperparameters of the above models have been tested proving that the above specified values are reasonable in terms of performance.
   In particular, when there is no limit on the depth of the decision tree, its performance is lower.  We have checked the reason for this and we found that this is  due to overfitting, 
   since the training performance was good, which is at odds with a low test performance. 
   %
   
\subsection{Performance estimates}\label{performanceestimates}
The performance metrics have been calculated comparing the expected labels and the predicted labels in the test set. This has been done with an iterative method, the 10 fold cross validation: the procedure divides the sample in ten parts after random shuffling. Based on this partition, it does ten iterations: in each of them nine parts of the ten folds construct the training set, and one is the testing set. As a consequence, a different fold represents the test set at each iteration. This procedure is the standard method used in statistics to understand how accurately a model will perform with new data. 
We highlight that this method  does not repeat the random definition of the ten folds in each iteration. Indeed, it is fundamental for this method to be effective to guarantee that the testing is done on a different fold in each iteration (as explained in the original paper by \citealt{stone1974}).
The testing set has been used for calculating the performance in each iteration. The 10 estimates of the performance metrics (that are  described below) have finally been averaged. We also checked that the performance in the ten iterations is not characterised by a large variance, but shows only marginal variations.

The samples used for the AGN-galaxies classifier and for type 1/2 AGN classifier are imbalanced (more AGN than galaxies and more type 1 than type 2), which might negatively affect the performance of the minority classes represented by galaxies and type 2 AGN.
To mitigate this effect, we used  the SMOTE (Synthetic Minority Oversampling Technique) algorithm  to oversample the minority class in the training set as was done in \cite{farrell2015}. The SMOTE method was independently applied fold by fold to increase only the training part, not the testing part of the sample, following the standard procedure which consists of creating extra training data \citep{chawla, ha1997}. 
We oversampled the minority classes up to become 50 per cent of the majority class in the training sample. 
SMOTE produces synthetic data from a real datapoint considering its 5 nearest neighbors, using the algorithm presented in \cite{chawla}:  
for each datapoint selects one of its five nearest neighbours at random.
It determines the vector that represents the distance between them (in the features space) and multiplies it by a random number between 0 and 1. In this way the new synthetic datapoint will be in one point of the distance vector between the original datapoint and its randomly selected neighbour.

The performance has been estimated starting from the confusion matrix (CM), which is
a simple but complete way to describe the performance of a classification model.
For a general class A - class B identification, CM is defined as:
\begin{equation*}
  CM = \begin{pmatrix}
    T_{A} & Mis_{A} \\
    Mis_{B} & T_{B}
    \end{pmatrix}
  \end{equation*}
Where
$T_A$ is the number of true objects of class A identified as A; $T_B$ is the number of B objects classified as B; $Mis_{B}$ is the number of objects belonging to class B but misclassified (classified as type A objects); $Mis_{A}$ is the number of objects  of class A misclassified (classified as B).
In the AGN-galaxy classification algorithm the CM contains: number of true galaxies $T_{g}$, number of true AGN $T_{AGN}$, number of false galaxies (in reality AGN) $Mis_{AGN}$, number of false AGN (in reality galaxies) $Mis_{g}$. 
For the type 1/2 classifier the CM contains: number of true type 1 AGN (in the matrix it is called $T_{type1}$); number of true type 2 ($T_{type2}$); number of misclassified type 2 objects, i.e. classified as type 1  ($Mis_{type2}$), number of misclassified type 1 objects, i.e. classified as type 2 ($Mis_{type1}$). The CM in the last case is therefore:
\begin{equation*}
  CM = \begin{pmatrix}
    T_{type1} & Mis_{type1} \\
    Mis_{type2} & T_{type2}
    \end{pmatrix}
  \end{equation*}

From the CM, we calculated the following metrics.
\begin{itemize}
\item
The precision of each class (AGN or galaxy in the first classifier, type 1  or type 2 in the second classifier), for example, the precision for type 1 AGN is: \[ P(type1)=  \frac{T_{type1}} {(T_{type1}+ Mis_{type2})}\]
As can be seen, the precision is best when minimising the number of false identifications of a given class.
\item The recall is a measurement of how many objects have been found of a class over the total number of examples of that class, for instance, for type 1 AGN: \[R(type1)=  \frac {T_{type1}} {(T_{type1} + Mis_{type1})}\]
 The recall is instead maximised by minimising the number of objects that belong to the class and have been missed by the classifier.
\item The F1 score is the harmonic average of precision and recall: \[F1 = \frac{(2*P*R)} {(P + R)}\] and it has been calculated for each class separately. This score for a perfect performance should be 1,  while 
it would be 0.5 for a random election
\item  The standard accuracy is: \[a = \left(\frac{T_{type1}+T_{type2}}{T_{type1}+Mis_{type1}+T_{type2}+Mis_{type2}}\right)    \]
\item Another performance metric investigated in this paper is the balanced accuracy, defined as the average of the recalls, for instance for the type 1/2 classification:
\[ba =  \frac{1}{2} \left(R_{type1}+ R_{type2}\right)    \]
\end{itemize}

The balanced accuracy is suitable for testing a binary classifier where the performance is different for the two classes. Its value for perfect predictions is 1, for random prediction is 0.
There are cases where the classifier performs differently for the two classes but the standard accuracy is good thanks to the imbalanced test set, in those cases the balanced accuracy is preferred.

\subsection{Features importance}\label{featimportance}

  The relevance of the features has been studied using the technique of permutation feature importance \citep{breiman2001}. This method 
  starts by computing the performance of the algorithm, after that it starts to analyse the features one by one. For each feature, it randomises its values in the dataset column producing a perturbed dataset. It repeats this randomisation process several times and in each of them it calculates the performance of the algorithm. After having finished all iterations, the average performance among all of them is calculated. 
The importance of a feature is defined as the
difference between the averaged performance (obtained from the randomisation) and its initial value (from the real dataset).

In our setup, we do 30 iterations for each feature and we compute the algorithm performance using the precision; we applied this method to the fitted algorithms adopted in this work.

\section{Results and discussion}\label{discussion}
The selected features are given as input to the ML algorithms in order to predict labels in the two test cases: AGN-galaxies in the first case and  type 1/2 AGN in the second case. The performance of the different ML algorithms was estimated with the 10 fold cross validation for each of the two classes in both test cases.
The performance has been evaluated in terms of the metrics described in Section \ref{performanceestimates}. We also investigated the influence of instrumental features or
directly source-related
features to the model training. This was done
to understand if 
such
parameters alone can be enough to identify the extragalactic sources in question. 
 We additionally tested the importance of spectroscopic redshifts among the features by checking the model performance after having replaced them with photometric redshifts; we finally made a test removing 
redshifts
during the model training.
\subsection{Classification AGN - galaxies}\label{discussionAGNgalaxies}
The first classification experiment aims at building a classifier based on the sample composed by 25599 AGN and 7262  galaxies. The performance metrics are discussed below. The analysis described below has also been repeated after relabeling 555 galaxies as AGN, since they have some evidence for AGN in the literature (as reported in the Milliquas survey), giving similar results. Another test has been made retraining the model after relabeling a broader sample of 4355 galaxies, based on their X-ray luminosity, higher than expected for an non-active galaxy; the results of this redefinition of the sample are discussed in Appendix \ref{relabelledAGNgalaxies}. 
To simplify
the discussion below, we refer to the performance metrics of the AB classifier with SMOTE
with spectroscopic redshifts included among the features
(shaded line of Table \ref{tab:resAGNgal}), unless differently stated. We note that all ML algorithms trained with spectroscopic redshifts give very similar results, therefore any of them could be used as a reference and would lead to the same conclusions.

\subsubsection{Performance metrics}

  The performance metrics in terms of precision, recall and F1 score,
  are reported in Table \ref{tab:resAGNgal}. 
  The classifiers perform better for AGN than for galaxies; removing redshifts degrades the performance of the classifiers, especially for galaxies. This can be intuitively understood as an effect of sample imbalance and of a better sampling of the AGN class in redshift; besides that, the X-ray spectra of AGN come from well known radiative processes in the nucleus itself, while for galaxies there can be radiation of different nature (binaries or star formation), which would make their X-ray spectra a mix of these contributions.
  More details on these aspects are discussed below.
  
  The algorithms reach a good performance and changing their hyperparameters does not
  impact the results: changing the number of SMOTE
  nearest neighbours from 5 to 10 has no impact as well as increasing
  the SMOTE ratio between the minority and the majority class to 0.9.  We also notice that the
  performance is rather similar for the different ML algorithms
  used. With redshifts included among the features, our results show
  that the precision in identifying AGN is $\sim95$ per cent while for
  galaxies it reaches 77 per cent (see shaded line of Table \ref{tab:resAGNgal}).  The identification for AGN reaches
  recall as high as 93 per cent, while the galaxies recall is $\sim$82
  per cent. 
  The classifiers reach such good
  results mainly thanks to X-ray data which provide the most effective way to find AGN: while X-ray radiation above a certain threshold constitutes evidence for an active nucleus, it allows to detect nuclei obscured in other wavelengths \citep{brandt2005}.
  The main limitation to the recall of 93 per cent could be the presence of sources with low X-ray luminosities, 
  but defined as AGN in the optical catalogues. There are 167 misclassified AGN in the test sample in Fig. \ref{flux_z_activity} and part of them cover a low flux - low redshift region of the plane in Fig. \ref{flux_z_activity}, since they are below the line of luminosity $10^{42}~\rm{erg~s^{-1}}$ (continuous line in the same scatter plot).
  Connected
  to this, there is a limitation in the precision of the galaxy
  classification, which does not reach levels higher than 78 per cent with the ML classifiers trained in this work (see Table \ref{tab:resAGNgal}). The precision of galaxies is limited due to the AGN classified as galaxies by the ML model.
  This is because part of sources labeled as AGN in the training sample (thus, in the optical spectroscopy) are
  not seen as AGN by the ML classifier. One of the likely reasons is that the X-ray features used to train the model are affected by absorption; as a consequence our ML model classifies them as galaxies, which gives a high rate of misclassified AGN.  However, heavily absorbed AGN are expected to be a small percentage in the survey, so other reasons such as intrinsically faint nuclei could bring the ML classifier to identify sources as galaxies, when they are optically classified as AGN.

  On the other hand, the limited recall of galaxies, which can reach a maximum 82 per cent level with our ML models,  by definition limited by the number of misclassified galaxies,  is  most probably due to the presence of sources in the training sample with galaxy labels from optical surveys, but with a conspicuous X-ray emission with  L$>10^{42} ~\rm{erg~s^{-1}}$. Fig. \ref{flux_z_activity} shows that almost all the 138 misclassified galaxies are above 
  that limit.
  As mentioned in Section \ref{type1type2trainingsample} there are 4355 sources in total with no evidence for nuclei reported in the optical surveys used in this work, but with high X-ray luminosities. High luminosities might cause the model to classify them as AGN, even though an AGN was not reported in the optical data. Relabeling these 4355 sources as AGN slightly improves the performance metrics in the AGN classification (see Appendix B), with an AGN precision and recall improvement of $\sim$3 per cent (e.g. see AB algorithm with SMOTE and redshifts included in Table \ref{tab:resAGNgal} and in Table \ref{res:enlargedAGNgal}). On the other hand, the galaxy classification from this test has a precision reduced by 12 per cent and a recall improved by 4 per cent (comparing the same lines of Table  \ref{tab:resAGNgal} and Table \ref{res:enlargedAGNgal}). We deduce that when relabeling these 4355 optical galaxies with high X-ray luminosity, the classifier is able to achieve a very good performance for AGN, but does not significantly improve the galaxy classification with a balanced accuracy improvement of only 3 per cent (from the same tables as before).
  More details of this test are shown in the Appendix B. 
  The F1 score of 94 per cent for AGN and 79 per cent for
  galaxies (see shaded line of Table \ref{tab:resAGNgal}) means that our classifier identifies both classes with a
  performance well above random (which would instead be characterised
  by a F1 score of 50 per cent). The F1 score also indicates a better
  performance for AGN than for galaxies, reflecting the performance
  metrics just discussed. There are multiple reasons for this: first,
  again the AGN oversampling; 
  second,
  galaxies are strongly limited in redshift.


  As explained above, the AGN-galaxy classification uses labels from optical spectroscopy, but part of the SDSS galaxies might have evidence for AGN outside SDSS. 
  We
  are not dealing here with the true intrinsic nature of the sources that we classify as AGN or galaxies, but we rather aim at providing labels as assigned from optical spectroscopic classifications.
  However, we explored the effect of mislabeling in a small
  percentage of the galaxy class since part
  of the galaxies ($\sim7$ per cent, from 555/7262 as mentioned in Section \ref{xmm-sdss-milli}) have some evidence for an AGN
  nucleus reported in the Milliquas survey. We repeated the analysis
  relabeling these sources as AGN and found very similar results,
  concluding that 7 per cent of mislabeled sources does not negatively
  affect the ML algorithms built with these large data sets.

\begin{table}[H]
  \begin{center}
    \caption{Performance metrics for AGN-galaxy classification with features set including spectroscopic redshifts (top table), photometric redshifts (central table), no redshifts (bottom table). SM indicates that the training sample was built with oversampling of the minority class (see text). Obs indicates that the training sample only includes observed data. $P_{AGN}$= precision calculated with respect to the AGN class.	 $P_{g}$= precision for galaxies class. $R_{AGN}$ = recall of AGN.  $R_{g}$ = recall of galaxies. $F1_{AGN}$ = F1 score of AGN. $F1_g$ = F1 score of galaxies. ba = balanced accuracy. RF, Tree, AB, grad, and vote are defined in Section \ref{algorithms}.}
    \label{tab:resAGNgal}
    Spectroscopic redshifts\\
\resizebox{.48\textwidth}{!}{    
    \begin{tabular}{l|c|c|c|c|c|c|c}
          \hline
algorithm	&	 $P_{AGN}$ 	&	 $P_{g}$ 	&	  $R_{AGN}$	&	 $R_{g}$ 	&	 $F1_{AGN}$ 	&	 $F1_g$   & ba\\
SM & & & & & & &  \\
RF	&	0.948	&	0.778	&	0.934	&	0.819	&	0.941	&	0.798	&	0.876 \\
Tree	&	0.95	&	0.737	&	0.916	&	0.83	&	0.933	&	0.781	&	0.873 \\
\rowcolor{LightCyan}
AB	&	0.947	&	0.77	&	0.931	&	0.817	&	0.939	&	0.793	&	0.874 \\
grad	&	0.947	&	0.777	&	0.933	&	0.816	&	0.94	&	0.796	&	0.875 \\
vote	&	0.949	&	0.776	&	0.932	&	0.825	&	0.941	&	0.800	&	0.879 \\
\hline
obs & & & & & & &  \\
RF	&	0.937	&	0.804	&	0.947	&	0.774	&	0.942	&	0.789	&	0.86 \\
Tree	&	0.936	&	0.772	&	0.935	&	0.776	&	0.936	&	0.774	&	0.855 \\
AB	&	0.934	&	0.807	&	0.948	&	0.762	&	0.941	&	0.784	&	0.855 \\
grad	&	0.935	&	0.816	&	0.951	&	0.766	&	0.943	&	0.790	&	0.858 \\
vote	&	0.937	&	0.815	&	0.95	&	0.773	&	0.943	&	0.793	&	0.862 \\
\hline
    \end{tabular}
}\\
  Photometric redshifts
\\
\resizebox{.48\textwidth}{!}{
    \begin{tabular}{l|c|c|c|c|c|c|c}
      \hline
 algorithm	&	 $P_{AGN}$ 	&	 $P_{g}$ 	&	  $R_{AGN}$	&	 $R_{g}$ 	&	 $F1_{AGN}$ 	&	 $F1_g$   & ba\\
SM & & & & & & &  \\
RF	&	0.961	&	0.844	&	0.937	&	0.892	&	0.949	&	0.865	&	0.915 \\
Tree	&	0.958	&	0.862	&	0.946	&	0.876	&	0.951	&	0.865	&	0.911 \\
AB	&	0.953	&	0.845	&	0.939	&	0.872	&	0.945	&	0.856	&	0.906 \\
grad	&	0.955	&	0.841	&	0.936	&	0.88	&	0.945	&	0.857	&	0.908 \\
vote	&	0.957	&	0.855	&	0.942	&	0.884	&	0.949	&	0.866	&	0.913 \\
\hline
obs & & & & & & &  \\
RF	&	0.959	&	0.858	&	0.945	&	0.885	&	0.951	&	0.869	&	0.915 \\
Tree	&	0.931	&	0.898	&	0.966	&	0.797	&	0.948	&	0.842	&	0.882 \\
AB	&	0.945	&	0.856	&	0.946	&	0.848	&	0.945	&	0.85	&	0.897 \\
grad	&	0.948	&	0.858	&	0.945	&	0.857	&	0.946	&	0.855	&	0.901 \\
vote	&	0.955	&	0.868	&	0.949	&	0.879	&	0.951	&	0.869	&	0.914 \\
\hline
    \end{tabular}
}\\
 No redshifts
\\
\resizebox{.48\textwidth}{!}{
    \begin{tabular}{l|c|c|c|c|c|c|c}
      \hline
 algorithm	&	 $P_{AGN}$ 	&	 $P_{g}$ 	&	  $R_{AGN}$	&	 $R_{g}$ 	&	 $F1_{AGN}$ 	&	 $F1_g$   & ba\\
SM & & & & & & &  \\
RF	&	0.848	&	0.625	&	0.93	&	0.412	&	0.887	&	0.496	&	0.671 \\
Tree	&	0.849	&	0.579	&	0.911	&	0.427	&	0.879	&	0.491	&	0.669 \\
AB	&	0.845	&	0.644	&	0.938	&	0.396	&	0.889	&	0.490	&	0.667 \\
grad	&	0.846	&	0.646	&	0.938	&	0.399	&	0.89	&	0.493	&	0.668 \\
vote	&	0.847	&	0.66	&	0.941	&	0.401	&	0.892	&	0.499	&	0.671 \\
\hline
obs & & & & & & &  \\
RF	&	0.839	&	0.717	&	0.961	&	0.349	&	0.896	&	0.469	&	0.655 \\
Tree	&	0.837	&	0.671	&	0.952	&	0.348	&	0.891	&	0.458	&	0.65 \\
AB	&	0.83	&	0.775	&	0.976	&	0.295	&	0.897	&	0.428	&	0.635 \\
grad	&	0.832	&	0.762	&	0.972	&	0.309	&	0.897	&	0.440	&	0.641 \\
vote	&	0.837	&	0.752	&	0.969	&	0.333	&	0.898	&	0.462	&	0.651 \\
    \end{tabular}
    }
  \end{center}
\end{table}

\subsubsection{Importance of features}

  The importance of features has been investigated with the permutation method described in Section \ref{featimportance}. We refer to the results of the permutation with the AB classifier, as everywhere else in this section, to simplify our discussion. We note indeed that all algorithms give similar permutation feature importance.  
  These results are reported in Table \ref{AGNgal_permphysfeat}. The permutation shows that redshifts represent the most relevant feature, which is most likely due to the prevalence of galaxies at lower redshifts and the prevalence of AGN at higher redshifts.
The permutation importance of the other features is at most 0.03; this means that the reference performance metrics change less than $\sim$3 per cent when the given feature is randomised. 
The redshift distributions will be discussed in more detail in Section \ref{zfluxAGNgal}.

\begin{table}
  \caption{Permutation feature importance of the AGN-galaxy classifier using the AB algorithm trained with SMOTE. Importance: mean of importances from iterations,  STD: standard deviation of the importances.}
  \label{AGNgal_permphysfeat}
  \begin{tabular}{l|c|c}
    Feature & importance & STD \\
    \hline
Z    &   0.216 &  0.004\\
$SC_{-}DET_{-}ML$ &  0.032 & 0.002\\
$SC_{-}EXT_{-}ML$ &  0.010 & 0.001\\
$SC_{-}EXTENT$ &  0.009 & 0.001\\
$EP_{-}ONTIME$  &  0.008 & 0.002\\
 $EP_{-}8_{-}CTS$ &  0.005  & 0.001\\
$EP_{-}OFFAX$    &  0.004 & 0.001\\
$DIST_{-}NN$ & 0.001 & 0.001\\
$N_{-}DETECTIONS$ &  0.001   & 0.001\\
$SUM_{-}FLAG$ &    0.001 & 0.001\\
$EP_{-}FLUX_{-}2_{-}3_{-}4$  &   0.000 & 0.000\\
 $EP_{-}8_{-}FLUX$ & 0.000   & 0.000\\
$EP_{-}5_{-}FLUX$   & 0.000 & 0.000
  \end{tabular}
\end{table}

We further investigated the importance of features with an additional test described in Appendix \ref{physfeatimportance}. This test demonstrated that the ML algorithms trained with source - related features only can achieve a similar performance as using all features in Table \ref{tab:featuresdescription}.
  \subsubsection{Impact of redshifts -  X-ray fluxes}\label{zfluxAGNgal}
  We have investigated how selection effects might affect the
  classification algorithms. We mainly explore flux and redshift parameters, since galaxies are preferentially found at
  low redshifts, while the AGN distribution is extended up to higher
  redshifts, with consequences in the classification performance when the data set is used to train the ML models.
  \begin{figure*}
    \includegraphics[width=8cm, height=6cm]{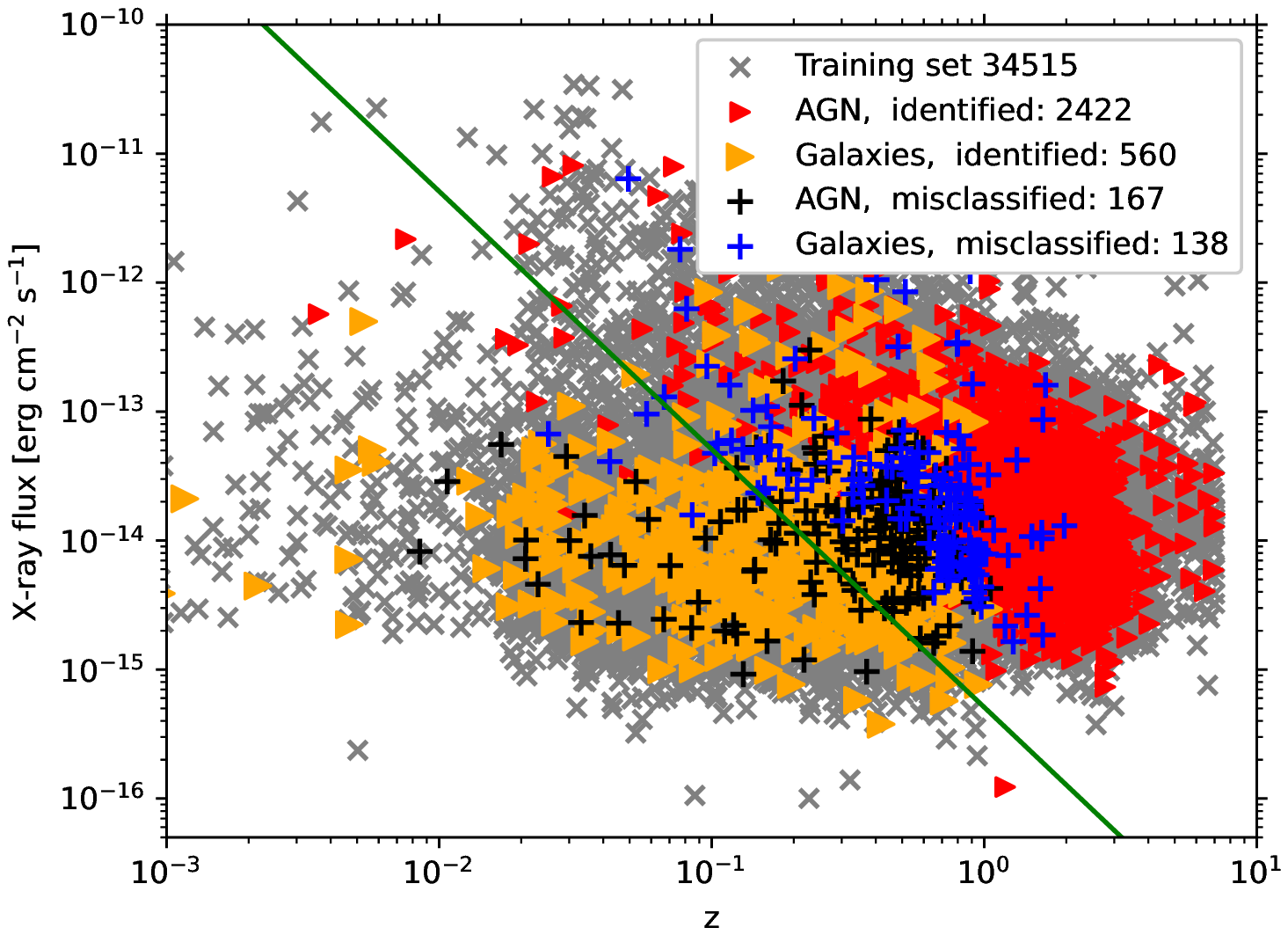}
     \includegraphics[width=8cm, height=6cm]{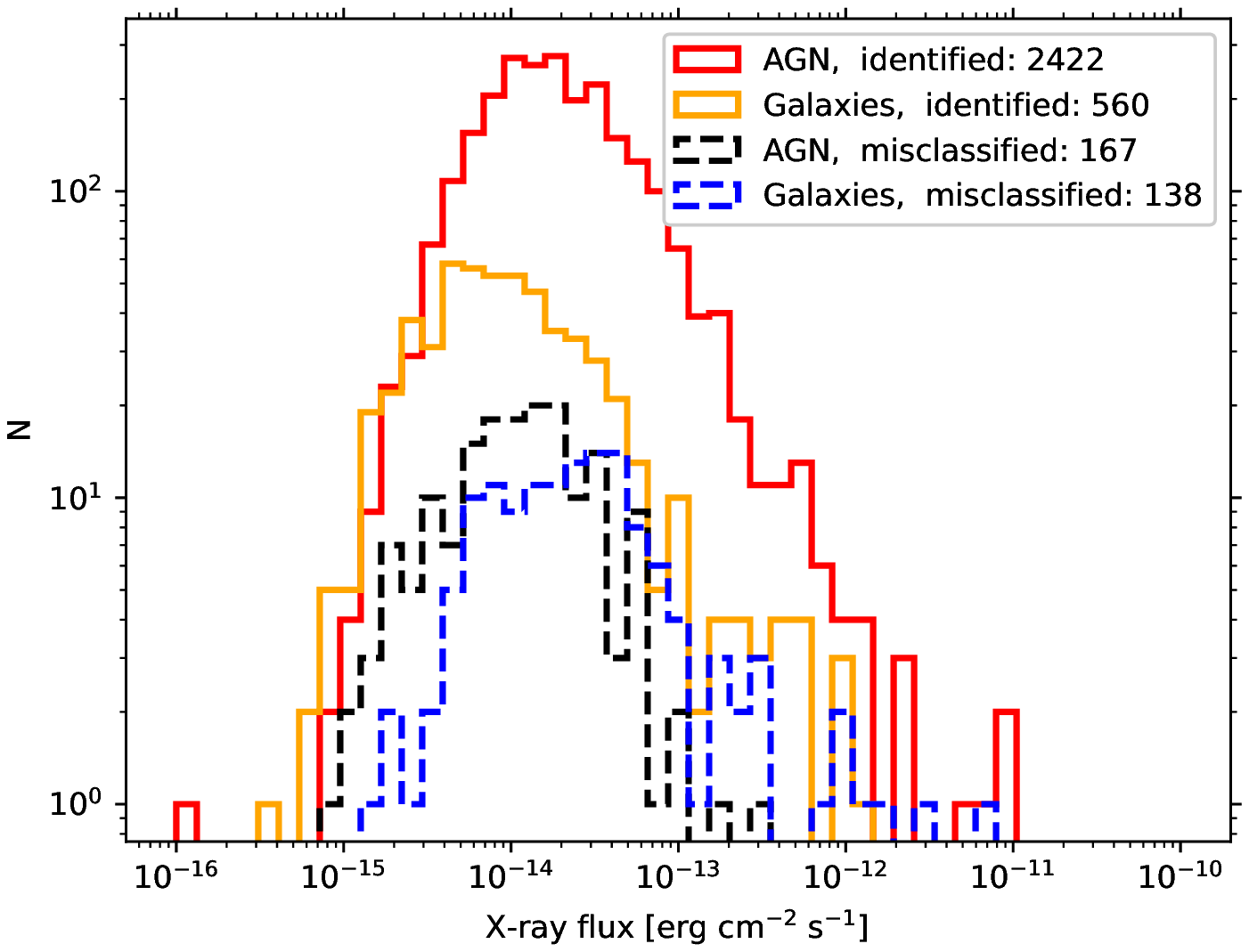}   \\
     \includegraphics[width=8cm, height=6cm]{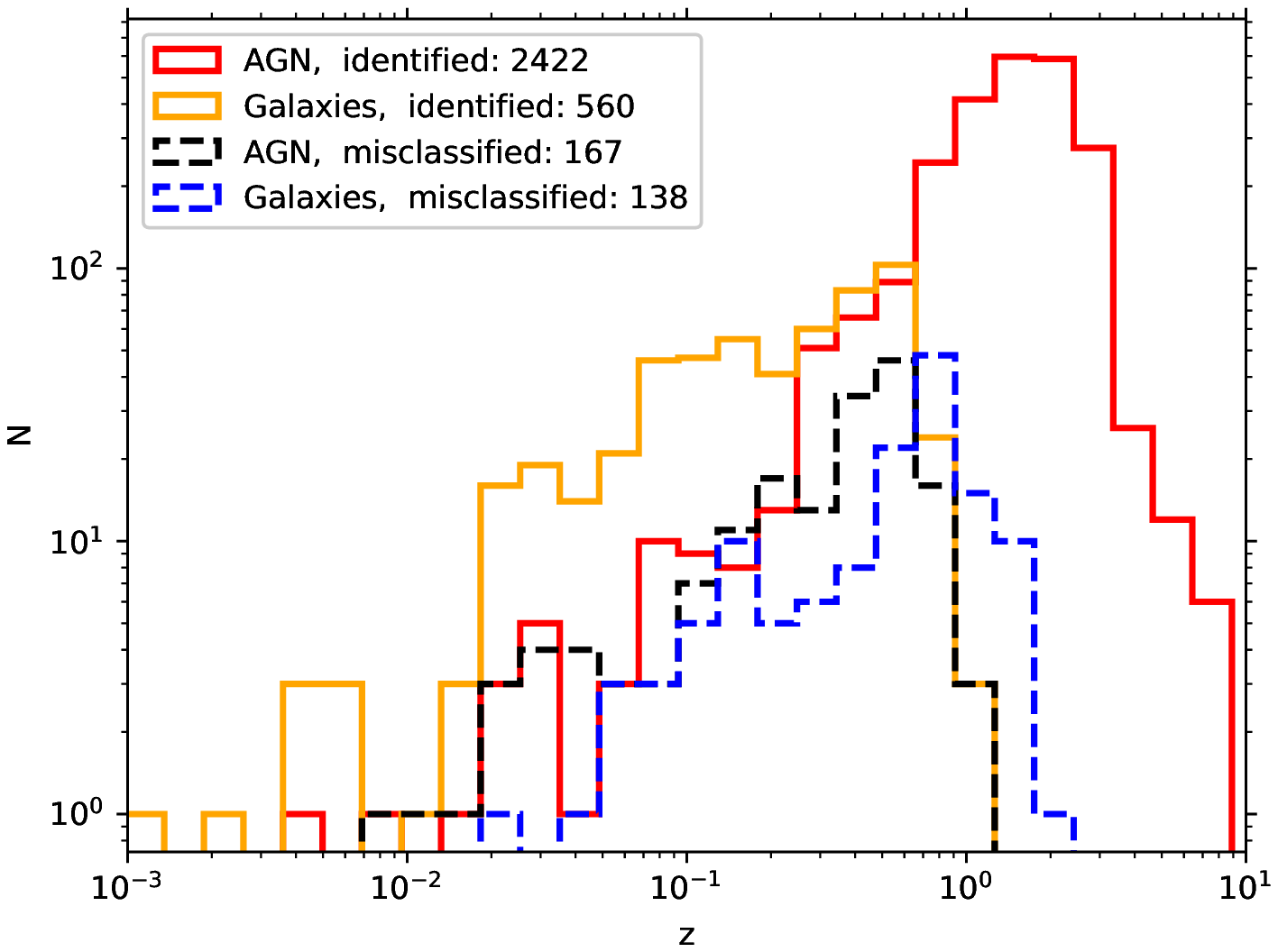}
     \includegraphics[width=8cm, height=6cm]{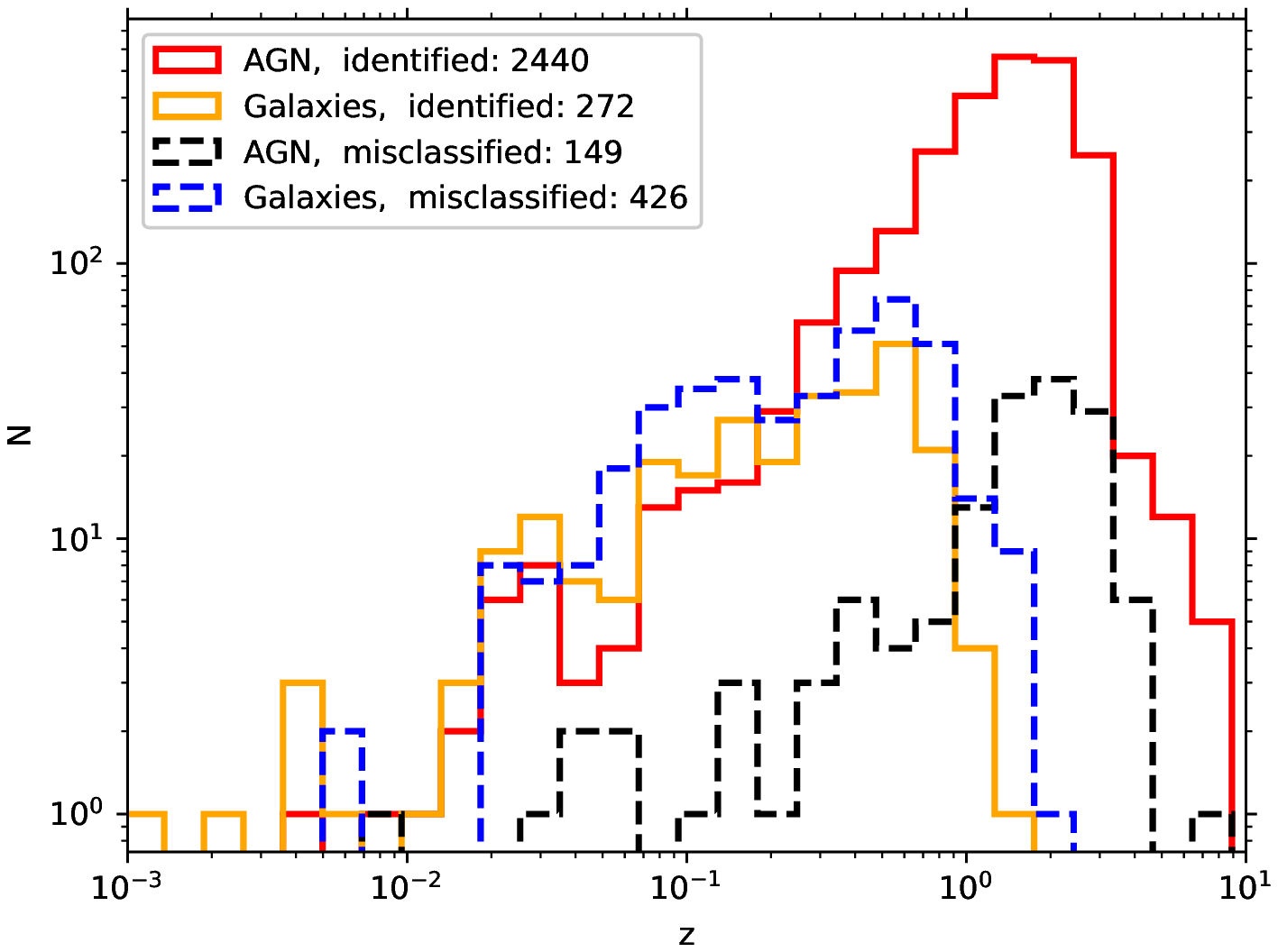}  
     \caption{Distribution of a test sample (with the corresponding training sample in the background) in redshift and X-ray flux (in band 2+3+4, see text), from the AGN-galaxy classification. The model used to obtain the predictions has been obtained from one iteration of the cross validation with redshifts (top panels and bottom left panel) and from another iteration without using redshifts (bottom right panel). The training has been done with the AB algorithm with SMOTE. The green continuous line in the top left panel corresponds to the luminosity $10^{42} ~ \rm {erg ~s^{-1}}$.               }
         \label{flux_z_activity}
   \end{figure*}
  Fig. \ref{flux_z_activity} shows the distribution of a training
  and test sample splitting the dataset, as usual, into 90 per cent for the
  training set and 10 per cent for the testing set.  The figure shows that the redshift
  distribution of galaxies is extended up to redshift 1, while the AGN
  distribution peaks at redshift $\sim$2 and it is extended up to
  very high redshifts (z$\sim8$).

  Focusing on the galaxies (yellow triangles and
  blue crosses in Fig. \ref{flux_z_activity}), most of the misclassified
  galaxies (labeled as galaxies in the optical surveys but classified as AGN by the ML model) are at redshift z between 0.5 and 1, where the model
  assigns the AGN identification to sources that are galaxies
  in many cases. However, from Fig. \ref{flux_z_activity} we notice that most of the misclassified galaxies have luminosities above the threshold of $10^{42}~ \rm erg~s^{-1}$:  i.e., from Fig. \ref{flux_z_activity}, almost all 138 galaxies interpreted as AGN by our ML algorithm have luminosities beyond $10^{42}\rm{erg~s^{-1}}$ (continuous line in  Fig. \ref{flux_z_activity}). 
Moreover, many of these sources are located in an area at $2\times10^{-14}~ \rm  erg~cm^{-2}~s^{-1}$ and z$\sim0.9$, which corresponds to luminosities of 0.3$\times~10^{44}~ \rm erg~s^{-1}$.
  These luminosities are too high to be emitted from a galaxy without an active nucleus. 
  The histograms in Fig. \ref{flux_z_activity} show that, above redshift $\sim$0.7, the number of misclassified galaxies
  exceeds the distribution of the correctly identified
  ones (see blue histogram in Fig. \ref{flux_z_activity}).  The
  distribution of galaxies in X-ray flux shows identified galaxies (true positive galaxies) well
  above the misclassified galaxies at low fluxes, while the
  misclassified galaxies become more important at higher fluxes (larger
  than $10^{-13}~ \rm erg~cm^{-2}~s^{-1}$). This seems to suggest that high
  fluxes are interpreted by the classifier as an evidence for an AGN
  nucleus, which is somehow expected.

  The AGN classification achieves very good
  performance at all fluxes explored in this survey.   The identified AGN distribution (red line) is above the misclassified AGN
  distribution (black dashed line) at redshift beyond
  $0.2$. Misclassified AGN are instead concentrated at redshifts lower than 0.2,
  most likely because this is where most galaxies are found.

  We also made an experiment to quantify the performance achievable without spectroscopic redshifts. With this purpose, we defined a training sample that includes only sources with photometric redshifts from KiDS (see details in Section \ref{data}). The resulting sample is composed by 260 galaxies and 710 AGN; for the test we  substituted spectroscopic with photometric redshifts. This training experiment gave excellent results in term of performance, considering the performance metrics shown in Table  \ref{tab:resAGNgal} (middle table). Precision, recall and F1 score are similar for AGN and galaxies with spectroscopic or photometric redshifts (looking as usual at the AB classifier with SMOTE).

  The performance metrics
  degrade when redshifts are not included among the features. The
  distribution of the test sample made without using redshifts among
  the features is shown in the bottom-right panel of Fig. \ref{flux_z_activity}. The detection of AGN
  becomes more challenging for redshift higher than 1, since the misclassified
  AGN peak at redshift $\sim2$. The model trained without redshift does not learn selection effects which cause high redshifts to be populated preferently by AGN. The AGN recall is unchanged for AGN because on average the model is able to find the same rate of AGN.
 Although we found a dramatic decrease of performance for galaxies when training the ML algorithms  without any redshifts, the ability to classify AGN is marginally affected.
  This shows that including redshifts among the features is necessary to achieve a good classification performance of galaxies,
  due to their limited redshift span (and corresponding
  lack of statistics above redshift 1);
 the performance of the AGN classification is not strongly affected by redshifts since AGN have better statistics and 
  a broader redshift coverage.

  Other parameters besides fluxes and redshifts among the list of selected features do not give new information in addition to what has been discussed in this section.
Source extent $SC_{-}EXTENT$ is strongly overlapping for AGN and galaxies. AGN correctly identified are composed by a vast majority of pointlike sources  (only 0.4 per cent of them have extent larger than zero as mentioned in the previous section), but galaxies correctly identified are also dominated by pointlike sources in the X-rays (since only the most nearby sources would be resolved by \emph{XMM-Newton}), even though the fraction of extended sources among the galaxies becomes 26 per cent.
  Distributions in the remaining features do not highlight substantial differences between AGN or galaxies surveys.

\subsubsection{Class imbalance}
The sample used for this test case is composed by AGN which populate
the survey 3.5 times more than the galaxies; we tried to mitigate this imbalance with the SMOTE method to oversample the training data.  Besides the different
statistics, the two classes have a different coverage in the
parameter space, the minority sample composed by galaxies is limited
within redshift z$\leq1$, while the majority sample composed by AGN has a
broader redshift distribution.  
As
discussed above, the class imbalance affects mostly the ability of the
models to recognise the minority class. The improvement of the recall
for the minority class obtained with the SMOTE oversampling
(see Table \ref{tab:resAGNgal}) is only marginal, from 76 per cent in the real dataset to 82 per cent in the dataset including SMOTE (see Table \ref{tab:resAGNgal}, highlighted row). 
The simulations, while slightly improving the recall of galaxies, give a marginally lower precision for the same class, consequently the performance in terms of accuracy is similar. Oversampling does not affect the performance of AGN classification, since AGN are the majority class (see Table \ref{tab:resAGNgal}).

\subsection{Type 1/2 AGN classification}\label{resultsAGNtype}
In this section, we discuss the classifier obtained to distinguish between type 1 and type 2 AGN, this model has been trained over the sample of 24029 type 1 AGN and 5203 type 2 AGN. For the discussion below, we use AB algorithm with SMOTE with spectroscopic redshifts included among the features as a reference model (shaded line of Table \ref{tab:restype1type2}), unless differently stated. As in the test case 1, all algorithms trained with the same features give similar results, therefore our conclusions do not depend on the specific reference algorithm chosen for our discussion. As done in the previous test case, we made a consistency check to understand the classification performance with a features set reduced to describe 
directly source-related
parameters, therefore the model has been retrained with a reduced features set. The test has demonstrated the ability of the model to generalize without using the features related to the instrumental conditions of these specific observations.

\subsubsection{Performance metrics}
The performance metrics are reported in Table \ref{tab:restype1type2}, and have been estimated with the same procedure adopted for the previous test case.
The different ML algorithms have very similar performance when trained on the same dataset. On the other hand, the best performance is achieved when the classifiers are trained including  redshifts in the features.

Varying the algorithm hyperparameters or the SMOTE number of neighbors does not lead to significantly better results. Similarly, the impact of changing the SMOTE ratio is very small (as discussed in the next paragraph), so we continue to discuss the results with a 0.5 SMOTE ratio, as was done in the previous paragraph.

The performance of different algorithms is rather similar when  using the same features. The best performance is achieved when the features include redshifts, with very good  results in the identification of type 1 AGN. When the training uses spectroscopic redshifts, the classifiers reach levels of precision of  96 per cent and a recall of 94 per cent for the type 1 class (see shaded line of Table \ref{tab:restype1type2}). For the more challenging task of identifying type 2 AGN, the precision is 74 per cent with a recall of 80 per cent for the reference model. The  F1 score is 95 per cent for type 1 AGN and 77 per cent for type 2 for this model.  This results in a balanced accuracy of 87 per cent,  which is a good result despite the severe selection effects (in redshifts and flux, as explained in Section \ref{zfluxtype1type2}) of the type 2 subsample.

When redshifts are excluded from the features, the performance drops for both classes, impacting in particular the recall of the minority class (type 2). In absence of redshifts, the ML algorithms tend to assign the type 1 label to sources that are instead type 2, with a very low recall of type 2 (26 per cent obtained with the AB algorithm with SMOTE, see Table \ref{tab:restype1type2}). 
This is due to the strong selection effects of the type 2 AGN resulting in a smaller redshift span, which limits their statistics beyond redshift 1.

\begin{table}[H]
  \begin{center}
    \caption{Performance metrics for type 1/2 classification with features selection including spectroscopic redshifts (top table), photometric redshifts (middle table), no redshifts (bottom table).
SM indicates that the training sample  has been built with oversampling of the minority class (see text). Obs indicates that the training sample includes observed data only. $P_{type1}$= precision calculated with respect to the type 1 AGN class.	 $P_{type2}$= precision for type 2 AGN class. $R_{type1}$ = recall of type 1 AGN class.  $R_{type2}$ = recall of type 2 AGN class. $F1_{type1}$ = F1 score of type 1. $F1_{type2}$ = F1 score of type 2.  ba = balanced accuracy. The ML algorithms are the same as in Table \ref{tab:resAGNgal}.  } 
    \label{tab:restype1type2}
Spectroscopic redshifts\\
    \resizebox{.48\textwidth}{!}{
    \begin{tabular}{l|c|c|c|c|c|c|c}
   
 algorithm 	&	 $P_{type1}$ 	&	 $P_{type2}$ 	&	  $R_{type1}$	&	 $R_{type2}$ 	&	 $F1_{type1}$ 	&	 $F1_{type2}$   & ba\\
SM & & & & & & &  \\
      RF	&	0.955	&	0.754	&	0.944	&	0.795	&	0.949	&	0.774	&	0.87 \\
Tree	&	0.957	&	0.699	&	0.925	&	0.806	&	0.94	&	0.749	&	0.865 \\
\rowcolor{LightCyan}
AB	&	0.956	&	0.737	&	0.938	&	0.802	&	0.947	&	0.768	&	0.87 \\
grad	&	0.957	&	0.749	&	0.941	&	0.807	&	0.949	&	0.777	&	0.874 \\
vote	&	0.958	&	0.746	&	0.94	&	0.812	&	0.949	&	0.777	&	0.876 \\
\hline
obs & & & & & & &  \\
RF	&	0.943	&	0.807	&	0.962	&	0.734	&	0.953	&	0.768	&	0.848 \\
Tree	&	0.939	&	0.77	&	0.954	&	0.715	&	0.946	&	0.741	&	0.834 \\
AB	&	0.942	&	0.797	&	0.96	&	0.728	&	0.951	&	0.760	&	0.844 \\
grad	&	0.944	&	0.800	&	0.96	&	0.739	&	0.952	&	0.768	&	0.849 \\
vote	&	0.945	&	0.809	&	0.962	&	0.74	&	0.953	&	0.773	&	0.851 \\
\hline
    \end{tabular}
    }\\
Photometric redshifts\\
\resizebox{.48\textwidth}{!}{
   \begin{tabular}{l|c|c|c|c|c|c|c}
\hline
 algorithm	&	 $P_{type1}$ 	&	 $P_{type2}$ 	&	  $R_{type1}$	&	 $R_{type2}$ 	&	 $F1_{type1}$ 	&	 $F1_{type2}$   & ba\\
SM & & & & & & &  \\
RF	&	0.973	&	0.801	&	0.952	&	0.887	&	0.963	&	0.838	&	0.92 \\
Tree	&	0.958	&	0.824	&	0.963	&	0.816	&	0.96	&	0.815	&	0.889 \\
AB	&	0.965	&	0.794	&	0.949	&	0.857	&	0.956	&	0.818	&	0.903 \\
grad	&	0.97	&	0.797	&	0.951	&	0.885	&	0.96	&	0.832	&	0.918 \\
vote	&	0.973	&	0.828	&	0.959	&	0.889	&	0.966	&	0.855	&	0.924 \\
\hline
obs & & & & & & &  \\
RF	&	0.965	&	0.853	&	0.968	&	0.853	&	0.966	&	0.848	&	0.911 \\
Tree	&	0.952	&	0.803	&	0.956	&	0.795	&	0.954	&	0.793	&	0.876 \\
AB	&	0.961	&	0.837	&	0.964	&	0.841	&	0.962	&	0.831	&	0.903 \\
grad	&	0.962	&	0.802	&	0.954	&	0.846	&	0.957	&	0.815	&	0.9 \\
vote	&	0.968	&	0.835	&	0.963	&	0.867	&	0.965	&	0.842	&	0.915 \\
\hline
   \end{tabular}
}\\
No redshifts\\
\resizebox{.48\textwidth}{!}{
   \begin{tabular}{l|c|c|c|c|c|c|c}
\hline
algorithm 	&	 $P_{type1}$ 	&	 $P_{type2}$ 	&	  $R_{type1}$	&	 $R_{type2}$ 	&	 $F1_{type1}$ 	&	 $F1_{type2}$   & ba\\
SM & & & & & & &  \\

RF	&	0.863	&	0.598	&	0.957	&	0.298	&	0.907	&	0.398	&	0.627 \\
Tree	&	0.86	&	0.595	&	0.958	&	0.282	&	0.907	&	0.382	&	0.62 \\
AB	&	0.859	&	0.682	&	0.974	&	0.261	&	0.913	&	0.377	&	0.617 \\
grad	&	0.859	&	0.655	&	0.97	&	0.265	&	0.911	&	0.377	&	0.617 \\
vote	&	0.859	&	0.672	&	0.972	&	0.266	&	0.912	&	0.380	&	0.619 \\
\hline
obs & & & & & & &  \\
RF	&	0.856	&	0.793	&	0.987	&	0.234	&	0.917	&	0.361	&	0.61 \\
Tree	&	0.854	&	0.726	&	0.981	&	0.227	&	0.913	&	0.346	&	0.604 \\
AB	&	0.850	&	0.847	&	0.992	&	0.194	&	0.916	&	0.315	&	0.593 \\
grad	&	0.851	&	0.840	&	0.992	&	0.198	&	0.916	&	0.320	&	0.595 \\
vote	&	0.854	&	0.833	&	0.991	&	0.215	&	0.917	&	0.341	&	0.603 \\
   \end{tabular}
}\\
 \end{center}
\end{table}
\subsubsection{Importance of features}

  The  list of features was investigated with the permutation feature importance as we did in test case 1. The results are given in Table \ref{type1type2_permphysfeat}, and show that redshifts are the most relevant features. This is a consequence of the type 2 AGN dominating at low redshifts and type 1 AGN at higher redshifts, as will be discussed in more detail in Section \ref{zfluxtype1type2}. All other features result in a permutation importance at most 2 per cent, which is a similar to what we found in the test case 1. We have further investigated the impact of features with the same procedure adopted for test case 1. The test demonstrates that the performance of the ML classifier does not change when the training is done with a reduced set of features including only source - related parameters. More details are shown in the Appendix \ref{physfeatimportance}.

\begin{table}
  \caption{Permutation feature importance of the type 1/2 classifier trained using the AB algorithm with SMOTE. Importance: mean of importances from iterations,  STD: standard deviation of the importances.}
  \label{type1type2_permphysfeat}
  \begin{tabular}{l|c|c}
    Feature & importance & STD \\
    \hline
Z    &   0.214 & 0.007\\
$SC_{-}DET_{-}ML$   & 0.022 & 0.003\\
$EP_{-}ONTIME$  & 0.018 & 0.002\\
$SC_{-}EXT_{-}ML$&  0.016 & 0.002\\
$SC_{-}EXTENT$  & 0.012 & 0.001\\
$EP_{-}OFFAX$ & 0.007 & 0.002\\
$EP_{-}8_{-}CTS$ & 0.003 & 0.001\\
$DIST_{-}NN$ &  0.001 & 0.000\\
$ SUM_{-}FLAG$  & 0.000 & 0.001\\
$N_{-}DETECTIONS$  & 0.000 & 0.001\\
$EP_{-}FLUX_{-}2_{-}3_{-}4$ & 0.000 & 0.000\\
$EP_{-}5_{-}FLUX$ & 0.000 & 0.000\\

  \end{tabular}
\end{table}

\subsubsection{Impact of redshifts - X-ray fluxes}\label{zfluxtype1type2}
\begin{figure*}
    \includegraphics[width=8cm, height=6cm]{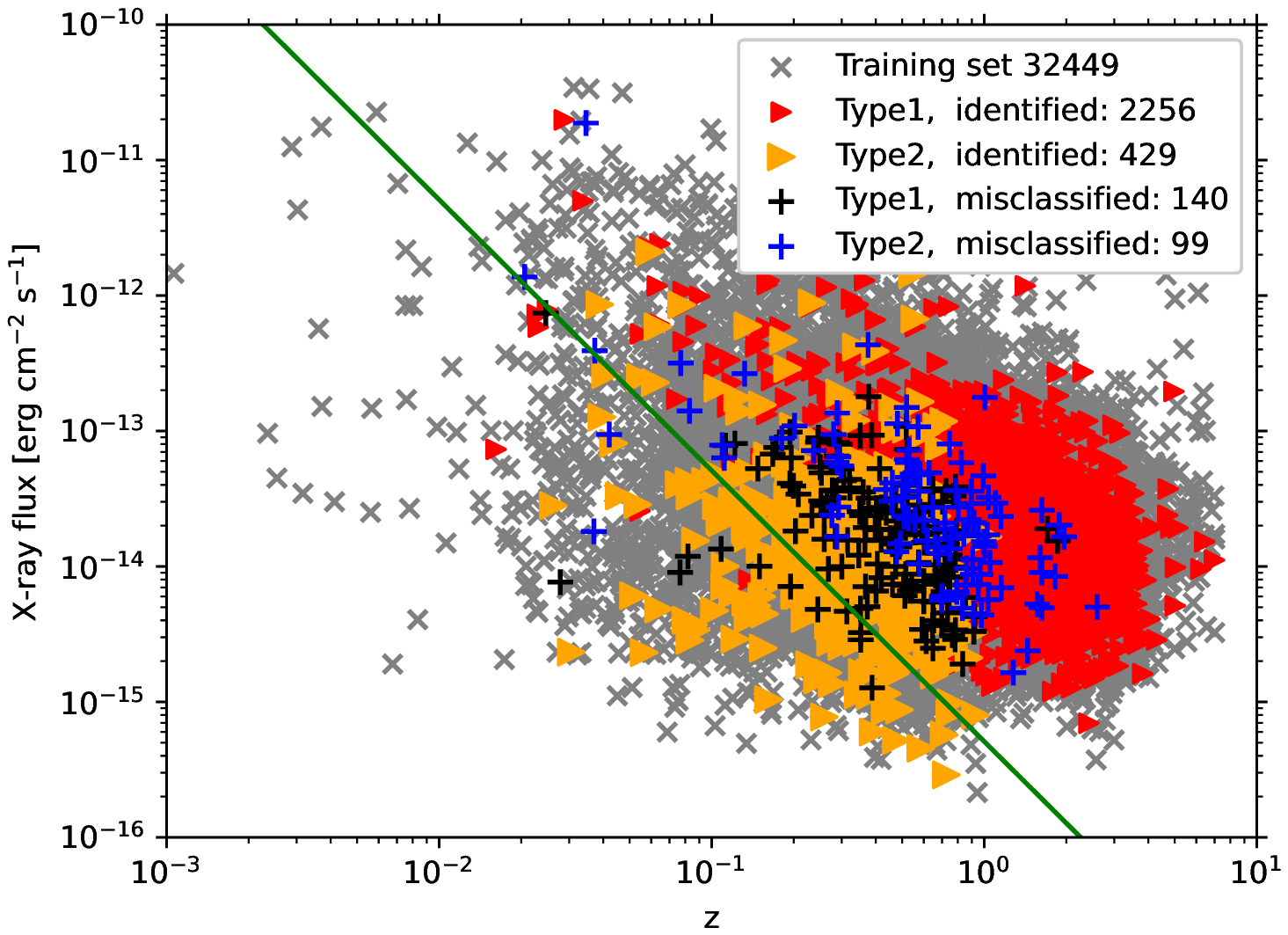}
     \includegraphics[width=8cm, height=6cm]{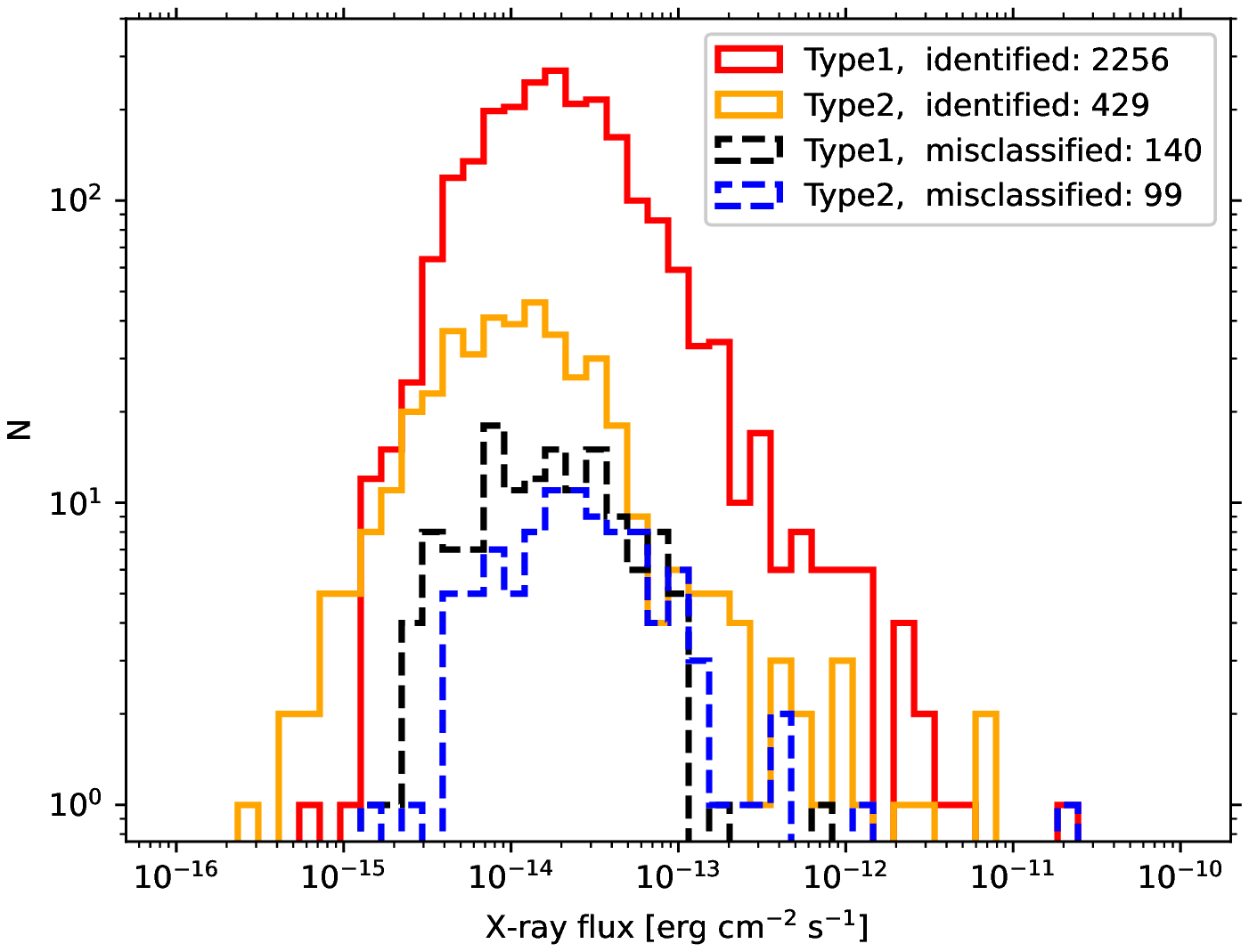}\\
     \includegraphics[width=8cm, height=6cm]{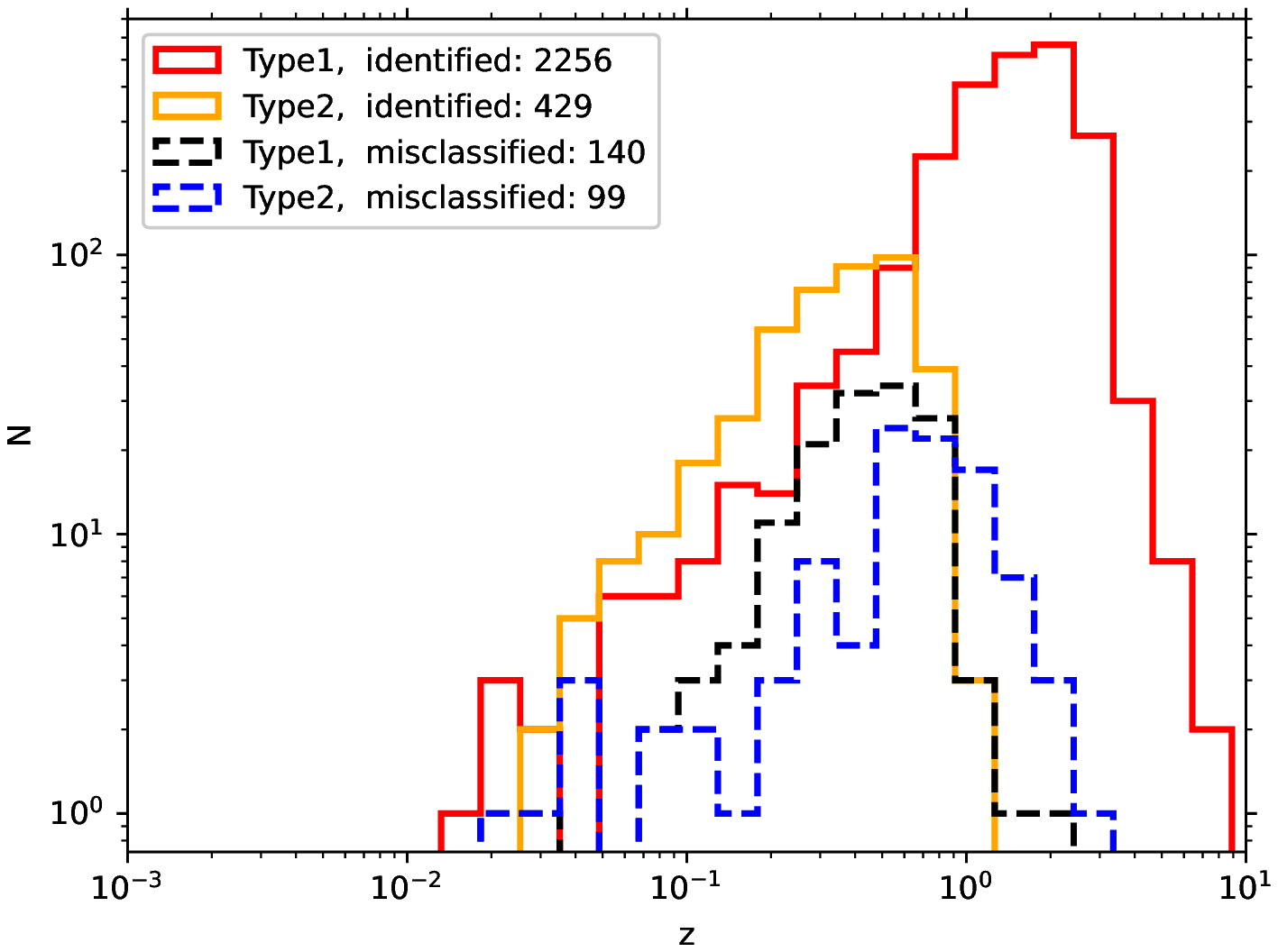}
     \includegraphics[width=8cm, height=6cm]{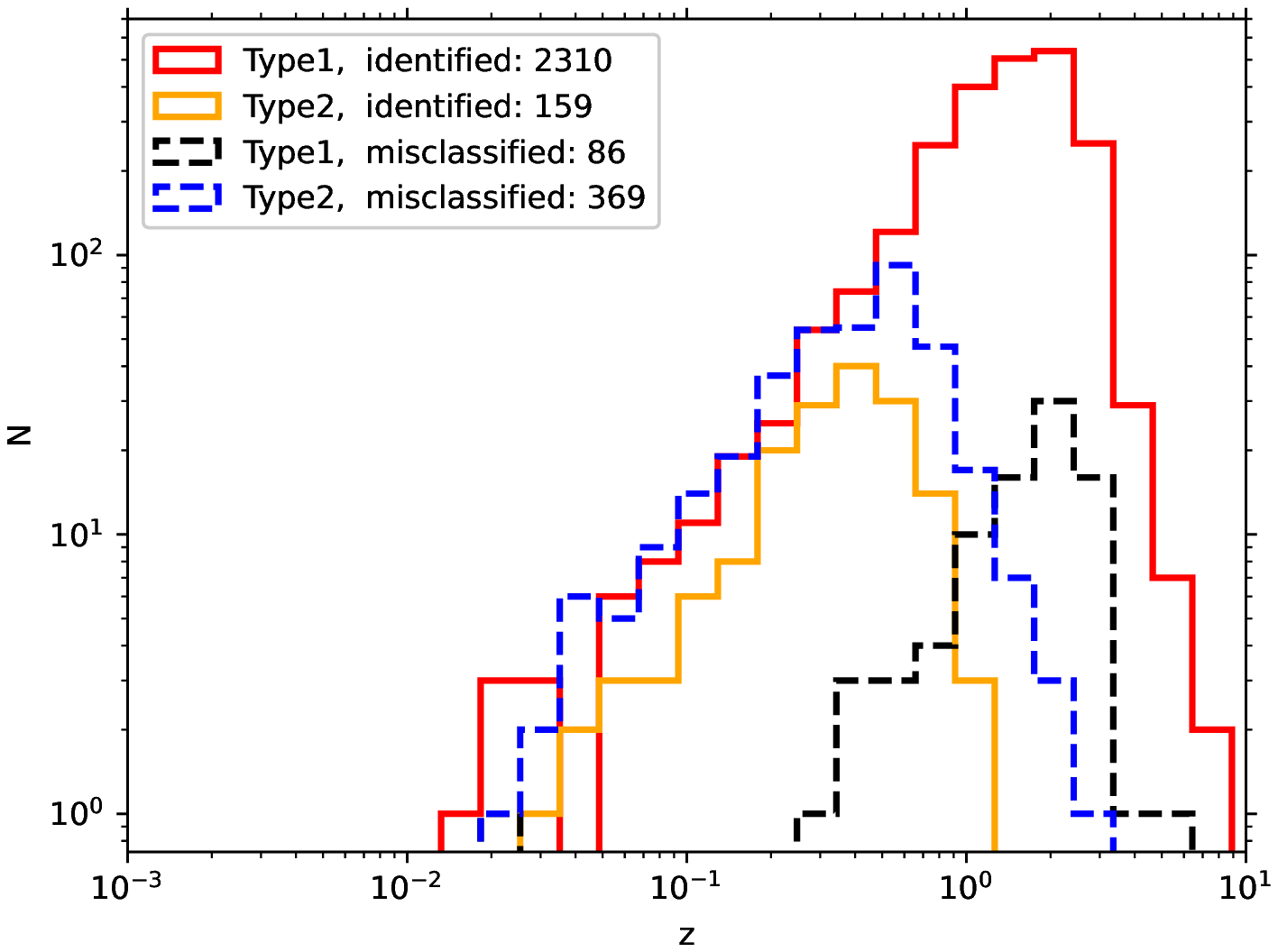}
      \caption{Distributions of test sample (and training sample in background) in redshift and X-ray flux (band 2+3+4) from the type 1/2 AGN classification. Top panels and bottom-left panel: distribution of the test sample (10 per cent of sample) with predictions from a model trained over 90 per cent of the sample using redshifts among the training features. Bottom right panel: results from a different training iteration made without including redshifts among the features. The training has been done with the AB algorithm with the SMOTE oversampling. The green continuous line in the top left panel corresponds to the luminosity $10^{42}~\rm{erg ~s^{-1}}$.   
              }
         \label{flux_z_AGNtype}
  \end{figure*}

The distribution of the sample in flux and redshift is
shown in Fig. \ref{flux_z_AGNtype}, which reports the test sample distribution resulting from one
iteration of the cross validation. The training has been done with SMOTE and training over all
features including redshifts in Fig. \ref{flux_z_AGNtype} except for the lower panel on the right (which has been trained without redshifts). The distribution of sources shows that
the misclassified type 2 AGN (which have type 2 labels but have been missed by our classifier) are mostly localised in a high density area at redshift
$\sim0.7-1$. A few sources at higher redshift z$>$1 have not been
labeled as type 2 AGN given the low occurrence of type 2 AGN in the
training dataset which reflects, more in general, their low number in the literature.  On the other hand, the classifier is successful for type 1 AGN, as can be seen in the
histograms in the top-right and bottom-left panel of Fig. \ref{flux_z_AGNtype}: the distribution of the identified type 1 AGN is above that of
the misclassified type 1 AGN at all redshifts and all fluxes.

We also performed an experiment to test the performance of the ML algorithms when trained with photometric redshifts instead of spectroscopic redshifts. This was done with the subsample of sources in common with the KiDS sample of photometric redshifts, resulting in 687 type 1 and 155 type 2 AGN. The ML algorithms were trained on this subset and using photometric redshifts instead of spectral ones, bringing good results in term of perfomance. As shown in Table \ref{tab:restype1type2}, the metrics are similar to those obtained for spectroscopic redshifts for both type 1 AGN and  type 2 AGN.

The performance of the ML algorithms decreases when redshifts are not included among features. The lack of redshift information affects the ability to select type 2 AGN, in particular in terms of recall (which decreases from 80\% to 26\% with the AB classifier, with SMOTE). 
The distribution of
type 1 and type 2 AGN in the model trained without redshifts is shown in the lower right panel of
Fig. \ref{flux_z_AGNtype}: the type 2 AGN classification model results in a
distribution of misclassified type 2 AGN above the classified type 2 at all
redshifts. The misclassified type 1 AGN are instead mostly located at redshift 1-1.2. It is interesting to notice that this is where the type 2 AGN distribution decays.

Besides redshifts and fluxes,  other features do not show a similarly clear difference in the distributions for type 1 and type 2 classes.

   \subsubsection{Class imbalance}
The training sample in each of the cross validation iterations has
$\sim$21600 (90 per cent of 24029) type 1 AGN and $\sim4670$ (90 per cent of 5203)
type 2 AGN. The imbalance ratio is $\sim$5:1 and we have tried to
mitigate its effects with SMOTE to oversample the training data. 
Running the ML algorithm with SMOTE 
improves the recall of the type 2 AGN, but at the same time decreases their precision. The overall accuracy of the  AB algorithm improves only by 3 per cent  introducing the SMOTE simulations (from 84 to 87 per cent comparing the AB line in SM and obs sections of Table \ref{tab:restype1type2}).
We tested if a different SMOTE ratio might improve the performance of the classifier.
A slight improvement in recall of the minority class of type 2 AGN was obtained by increasing the SMOTE ratio from 0.5 to 0.9, at the same time with a decrease in precision. This is a well known effect when increasing the size of a class with synthetic datapoints:  a lower number of false negatives with a corresponding  better recall is achieved (since the algorithms has been trained on a larger sample of that class), but the number of false positives is not guaranteed to decrease, hence not necessarily increasing the precision.

\subsection{Limitations of binary classification}
This work has presented two binary classifiers to study extragalactic surveys in order to distinguish AGN from galaxies (test case 1) and type 1 from type 2 AGN (test case 2), using X-ray features. 
Intermediate classes, including e.g. LLAGN, have not been taken into account to build this binary classifiers.

Regarding the first test case, even though a significant fraction of local galaxies shows some evidence of faint nuclei, they are not always detected in X-rays, and it is unfeasible to disentangle the nucleus from the host in the majority of the observations with the current X-ray instruments. X-ray spectra of galaxies shows a contribution from binaries, with typically hard spectra, which could actually be similar to the spectra of faint nuclei.  Literature reports that faint nuclei, when successfully disentangled from the light of the host, also show a hard powerlaw from the LLAGN (as explained invoking different scenarios, e.g. \citealt{connolly2016} or \citealt{ho2008} or \citealt{guainazzi2000}).

Another aspect which may affect the classification is  X-ray continuum variability. It has indeed been demonstrated  that variability features play a role in identifying AGN \citep{lo2014}. Variability could be found in virtually all AGN, from low to high luminosity, highlighting intrinsic mechanisms in the SMBH (for instance, \citealt{sobolewska2009,mchardy2006}).

While this aspect should not affect the ability to distinguish AGN and galaxies in the first test case, we believe it has an impact on the performance of type 1 AGN - type 2 AGN classification. From the Chandra Deep Field South survey, \cite{paolillo2004} reported a widespread intrinsic variability among AGN, prevalent in the ones unabsorbed in the X-rays. The authors found that the fraction of variable sources decreases with increasing intrinsic X-ray absorption. This kind of variability was explained by the authors as due to instabilities in the innermost regions of the nuclei.
A very different scenario is variability of column densities in AGN, which involves larger scales. Literature reports variable column densities in X-ray absorbed AGN \citep{risaliti2002} and extreme cases have been reported with
transitions from an absorbed to an unabsorbed phase.  For instance \cite{panessa2009} found a mismatch between the optical type 1 - type 2 classification and the X-ray absorption that was explained with a clumpy absorbing material in the torus. These extreme cases may at some extent complicate a type 1/2 classifier based on X-ray features when the labels of the training dataset are provided from optical spectra.

\section{Conclusions}\label{conclusions}

This paper tests several ML algorithms to classify extragalactic sources using samples based on XMM and SDSS.
The first classifier has been constructed to 
distinguish between 
AGN and galaxies and another classifier to divide the AGN class into type 1 (unabsorbed) and type 2 (absorbed). We trained the first classifier using a sample with 25599 AGN and 7262 galaxies. The model to divide AGN into type 1 and type 2 is trained over a data set which contains 24029 type 1 AGN and 5203 type 2 AGN (using Milliquasar data in addition to XMM and SDSS). The performance metrics were calculated with the cross validation, and the training data sets were oversampled such that the minority class constitutes 50 per cent of the majority class.
We tested several classifiers 
which reached accuracy levels of 87 per cent in both classification problems (balanced accuracy). 
The maximum difference of balanced accuracy between the ML algorithms adopted in this work is less than 1 per cent, demonstrating a similar performance. All of them are indeed based on decision trees, and converge to reach the best performance achievable with tree-based algorithms.   
Our conclusions discussed below refer to the performance metrics of the AB algorithm obtained with oversampling and including spectroscopic redshifts among the features, unless stated otherwise.

Good results are found when identifying AGN (94 per cent precision) in a sample of AGN and galaxies (first test case). A similar performance is reached when identifying type 1 AGN (96 per cent precision) in a sample of type 1 AGN and type 2 AGN (second test case).  The minority classes represent a major challenge, since the precision of identifying galaxies in the first test case is 77 per cent, similar to the precision of identifying type 2 AGN in the second test case, which reaches 74 per cent.  In the first test case, the precision of galaxies is limited due to a number of sources optically classified as AGN, but classified as galaxies by the ML algorithm: this is in part due to many of them having low redshift and low flux, which is a flux-redshift range where galaxies are preferentially found. These sources may be AGN that are intrinsically faint or absorbed in the X-ray band.
The performance in the second test case is limited by the decay of the type 2 AGN distribution at redshift 0.5 - 1 in the training sample: better statistics of type 2 AGN at such redshifts would help to improve the classification performance.

The ML algorithms were trained with spectroscopic and photometric redshifts, and we also performed a test without using redshifts among the features.
The test with photometric redshifts was done for the sources in common with KiDS (considering the surveys of bright galaxies and quasars). The performance was consistent with that obtained using spectroscopic redshifts.
By contrast, training the ML algorithms without redshifts, the performance is  drastically reduced for the minority classes (galaxies in the first test case and type 2 AGN in the second test case). We also noticed that the minority classes have a much more limited coverage of the flux-redshift space, due to stronger selection effects. We conclude that the main limitations are connected to the sparse sampling at high redshifts. 


Retraining the model with a reduced features set, which includes only
directly source-related
features has demonstrated the robustness of the results, highlighting the ability of our models to generalise well beyond the instrumental conditions of the training datasets.

Concluding, the classifiers trained with the current data can be applied in the future to identify AGN and galaxies as well as to distinguish between type 1 and type 2 AGN, for sources that already have some distance information. For new and unknown sources without such information, the current 
methods are suitable
to identify AGN in extragalactic surveys and to classify type 1 AGN  in AGN surveys; on the other hand, it cannot be successfully used alone in finding galaxies and type 2 AGN. Our expectations are that the current training sample should be complemented with future surveys with better statistics of galaxies without active nuclei and with an improved statistics of absorbed AGN:  retraining the current model would certainly help to achieve an acceptable accuracy for these two classes.

 Current surveys such as \textsl{eROSITA} or those planned to be carried out with the \textsl{Athena} mission will expand the known census of AGN by a factor 100, especially improving the demography of minority classes from low to high redshift \citep{nandra2013}. This will help to overcome the strong sensitivity limits of current surveys (including the training sample used in this work).  With an effective area of $\sim 1.4\ \rm{m}^2$ at 1 keV, the \textsl{Athena} survey will be sensitive enough to build  extra-galactic surveys with a variety of different classes, including heavily absorbed AGN up to $z\sim3$, and LLAGN at low redshifts. 
 
\section*{Acknowledgements}
The authors acknowledge the referee for insightful suggestions and the careful review of the paper.
FJC acknowledges financial support from
the Spanish Ministry MCIU under project RTI2018-096686-
B-C21 (MCIU/AEI/FEDER/UE), cofunded by FEDER
funds and from the Agencia Estatal de Investigación,
Unidad de Excelencia María de Maeztu, ref. MDM-2017-
0765.

Funding for the Sloan Digital Sky 
Survey IV has been provided by the 
Alfred P. Sloan Foundation, the U.S. 
Department of Energy Office of 
Science, and the Participating 
Institutions. 

SDSS-IV acknowledges support and 
resources from the Center for High 
Performance Computing  at the 
University of Utah. The SDSS 
website is www.sdss.org.

SDSS-IV is managed by the 
Astrophysical Research Consortium 
for the Participating Institutions 
of the SDSS Collaboration including 
the Brazilian Participation Group, 
the Carnegie Institution for Science, 
Carnegie Mellon University, Center for 
Astrophysics | Harvard \& 
Smithsonian, the Chilean Participation 
Group, the French Participation Group, 
Instituto de Astrof\'isica de 
Canarias, The Johns Hopkins 
University, Kavli Institute for the 
Physics and Mathematics of the 
Universe (IPMU) / University of 
Tokyo, the Korean Participation Group, 
Lawrence Berkeley National Laboratory, 
Leibniz Institut f\"ur Astrophysik 
Potsdam (AIP),  Max-Planck-Institut 
f\"ur Astronomie (MPIA Heidelberg), 
Max-Planck-Institut f\"ur 
Astrophysik (MPA Garching), 
Max-Planck-Institut f\"ur 
Extraterrestrische Physik (MPE), 
National Astronomical Observatories of 
China, New Mexico State University, 
New York University, University of 
Notre Dame, Observat\'ario 
Nacional / MCTI, The Ohio State 
University, Pennsylvania State 
University, Shanghai 
Astronomical Observatory, United 
Kingdom Participation Group, 
Universidad Nacional Aut\'onoma 
de M\'exico, University of Arizona, 
University of Colorado Boulder, 
University of Oxford, University of 
Portsmouth, University of Utah, 
University of Virginia, University 
of Washington, University of 
Wisconsin, Vanderbilt University, 
and Yale University.

Based on observations made with ESO Telescopes at the La Silla Paranal Observatory under programme IDs 177.A-3016, 177.A-3017, 177.A-3018 and 179.A-2004, and on data products produced by the KiDS consortium. The KiDS production team acknowledges support from: Deutsche Forschungsgemeinschaft, ERC, NOVA and NWO-M grants; Target; the University of Padova, and the University Federico II (Naples).

\section*{Data Availability}
The  surveys used to build the training samples are available in the archives.  The derived data
and models
will be shared upon request to the 
first author.



\bibliographystyle{mnras}
\bibliography{bibliography} 





\appendix

 \section{Details of sample selection}\label{appdatadetails}
\begin{figure}
   \includegraphics[width=9cm, height=8cm]{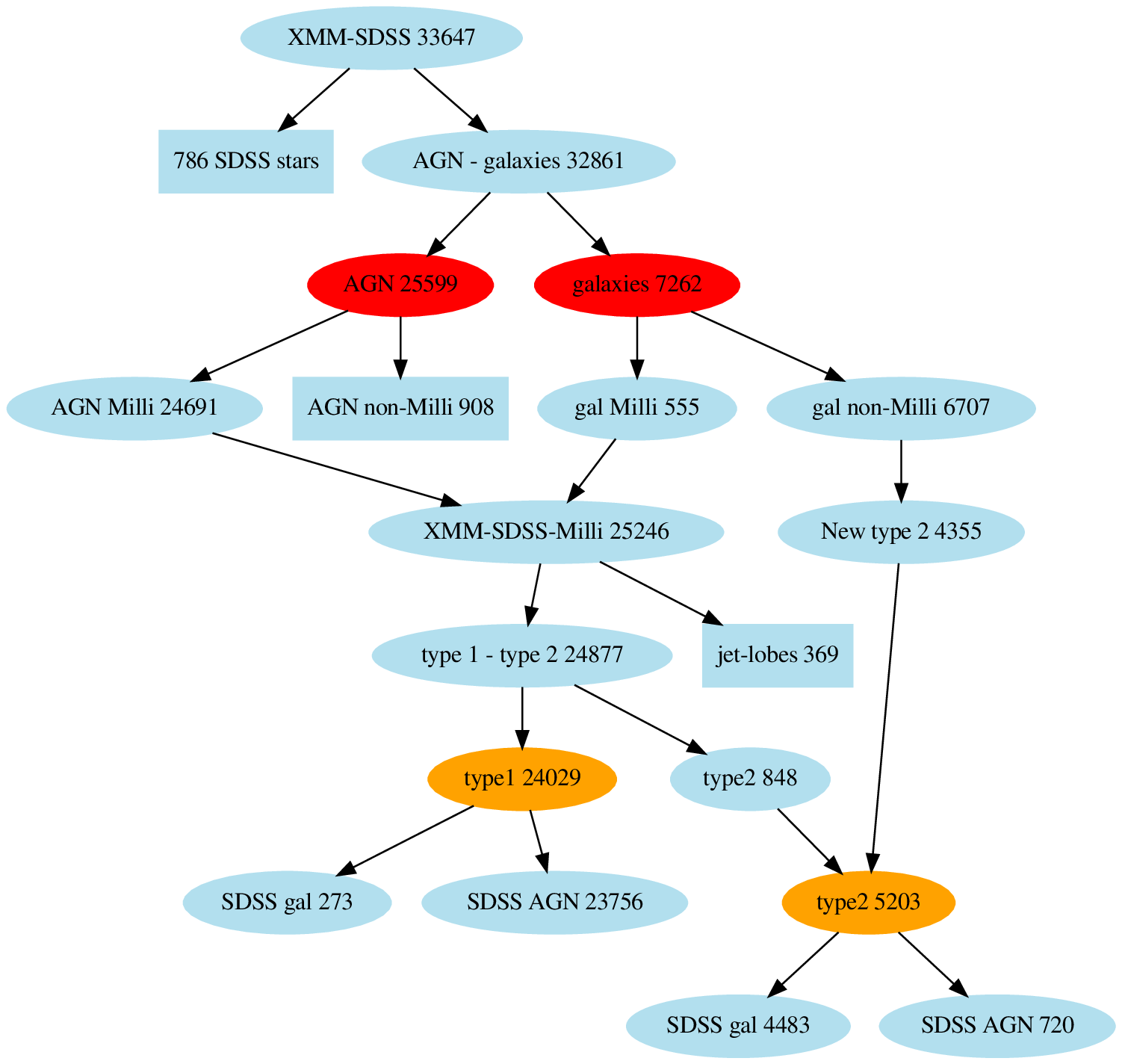}
      \caption{Sample composition in the two test cases: test case 1 with red background and test case 2 with orange background. In the fourth line of the diagram, Milli and non-Milli refer to sources included or not included in Milliquas, respectively.
              }
         \label{samplecomposition}
  \end{figure}

  The steps made to build the samples are shown in figure \ref{samplecomposition}, which indicate the number of sources of different classes.
  The test case 1, classifying AGN and galaxies, is done with the sample represented with red background in the figure. The test case 2 which classifies type 1 and type 2 AGN is instead based on the sample with orange background. The boxes  represent the sources excluded from this study, while the ovals show the different steps and the final samples. 
 \section{Expanding AGN sample with new type 2 candidates}\label{relabelledAGNgalaxies}
 We have redefined the training sample of the AGN-galaxy classifier taking into account the presence of the 4355 new type 2 AGN candidates defined in Section \ref{type1type2trainingsample}. Specifically, we switched the labels of these 4355 sources from galaxies to AGN to account for a possible obscured nucleus.
 The sample is therefore composed by 29954 AGN and 2907 galaxies.

 We then retrained the model to classify AGN and galaxies and the results are reported in Table \ref{res:enlargedAGNgal}. This model does not show a strong improvement in the galaxy classification, which raises its recall by only 3 per cent with respect to the previous training described in Section \ref{discussion}, focussing on our reference algorithm, AB with SMOTE and including redshifts. The improvement in the AGN classification is also around 3 per cent with the reference algorithm. Without including redshifts in the features, we notice that, while the AGN classification improves with this newly defined dataset, the galaxy classification degrades, especially without the SMOTE simulations, i.e. the recall is 7 per cent for the classifier without redshifts and trained with the observed sample only (see AB algorithm in last section of table \ref{res:enlargedAGNgal}). This strong degradation in galaxies recall clearly reflects the increased imbalance between galaxies (which result in a very poor performance) and AGN (with a very good performance), where galaxies represent only 9 per cent of the training sample. The SMOTE simulations improve the galaxy recall up to levels similar to those discussed in Section \ref{discussion}: the recall is around 35 per cent for the AB algorithm in the SMOTE section of Table \ref{res:enlargedAGNgal} without redshifts and 40 per cent in the same line of Table \ref{tab:resAGNgal}.

\begin{table}[H]
  \begin{center}
    \caption{Performance metrics for AGN-galaxy classification with features including source related properties and more instrumental related parameters. Spectroscopic redshifts are included in the top table, photometric redshifts in the middle table, no redshifts are included in  the bottom table. SM indicates that the  training sample has been built with oversampling of the minority class (see text). Obs indicates that the training sample only includes observed data. $P_{AGN}$= precision calculated with respect to the AGN class.	 $P_{g}$= precision for galaxies class. $R_{AGN}$ = recall of AGN.  $R_{g}$ = recall of galaxies. $F1_{AGN}$ = F1 score of AGN. $F1_g$ = F1 score of galaxies. ba = balanced accuracy.
 RF, Tree, AB, grad, and vote are defined in Section \ref{algorithms}.}
    \label{res:enlargedAGNgal}
    Spectroscopic redshifts\\
    \resizebox{.48\textwidth}{!}{  
    \begin{tabular}{l|c|c|c|c|c|c|c}
          \hline
algorithm 	&	 $P_{AGN}$ 	&	 $P_{g}$ 	&	  $R_{AGN}$	&	 $R_{g}$ 	&	 $F1_{AGN}$ 	&	 $F1_g$   & ba\\
SM & & & & & & &  \\
RF	&	0.983	&	0.733	&	0.971	&	0.832	&	0.977	&	0.778	&	0.901 \\
Tree	&	0.983	&	0.663	&	0.959	&	0.83	&	0.971	&	0.736	&	0.895 \\
\rowcolor{LightCyan}
AB	&	0.985	&	0.653	&	0.956	&	0.849	&	0.97	&	0.738	&	0.902 \\
grad	&	0.984	&	0.676	&	0.961	&	0.837	&	0.972	&	0.747	&	0.899 \\
vote	&	0.985	&	0.703	&	0.965	&	0.848	&	0.975	&	0.768	&	0.907 \\
\hline
obs & & & & & & &  \\
RF	&	0.976	&	0.838	&	0.986	&	0.751	&	0.981	&	0.791	&	0.868 \\
Tree	&	0.975	&	0.788	&	0.981	&	0.738	&	0.978	&	0.761	&	0.859 \\
AB	&	0.974	&	0.797	&	0.982	&	0.736	&	0.978	&	0.765	&	0.859 \\
grad	&	0.979	&	0.642	&	0.957	&	0.793	&	0.968	&	0.709	&	0.875 \\
vote	&	0.979	&	0.644	&	0.958	&	0.793	&	0.968	&	0.71	&	0.875 \\
\hline
    \end{tabular}
  }  \\
    No redshifts\\
    \resizebox{.48\textwidth}{!}{ 
    \begin{tabular}{l|c|c|c|c|c|c|c}
      \hline
 algorithm	&	 $P_{AGN}$ 	&	 $P_{g}$ 	&	  $R_{AGN}$	&	 $R_{g}$ 	&	 $F1_{AGN}$ 	&	 $F1_g$   & ba\\
SM & & & & & & &  \\
RF	&	0.935	&	0.353	&	0.942	&	0.325	&	0.939	&	0.338	&	0.634 \\
Tree	&	0.938	&	0.292	&	0.91	&	0.381	&	0.924	&	0.330	&	0.646 \\
AB	&	0.937	&	0.34	&	0.934	&	0.350	&	0.935	&	0.344	&	0.642 \\
grad	&	0.935	&	0.337	&	0.936	&	0.337	&	0.936	&	0.336	&	0.636 \\
vote	&	0.937	&	0.365	&	0.942	&	0.346	&	0.935	&	0.355	&	0.644 \\
\hline
obs & & & & & & &  \\
RF	&	0.92	&	0.517	&	0.99	&	0.115	&	0.954	&	0.187	&	0.552 \\
Tree	&	0.922	&	0.407	&	0.979	&	0.148	&	0.95	&	0.216	&	0.564 \\
AB	&	0.916	&	0.555	&	0.995	&	0.065	&	0.954	&	0.116	&	0.530 \\
grad	&	0.918	&	0.550	&	0.993	&	0.083	&	0.954	&	0.143	&	0.538 \\
vote	&	0.918	&	0.550	&	0.993	&	0.089	&	0.954	&	0.147	&	0.540 \\
\hline
    \end{tabular}
    }
    \end{center}
\end{table}

\section{Impact of features selection}\label{physfeatimportance}
 In this section we further investigate the impact of features in the performance of the ML algorithms in both test cases.
Given that the permutation method does not highlight the importance of the features beyond the first most relevant one (redshifts), we explore this in more detail with an additional experiment which consists of reducing the list of features. 
The models constructed for the AGN-galaxy classification problem were initially trained using the 13 features listed in Table \ref{tab:featuresdescription}. 
 The performance of the model has been tested by removing all instrumental features in order to see their genuine importance. We therefore trained a new model with the
directly source-related
features only and checked its performance metrics. The results, shown in Table \ref{tab:resAGNgalphys}, are very similar to the ones obtained with the larger list of features (reported in Table \ref{tab:resAGNgal}). This demonstrates that a good model training can be achieved with source-related features only, since the others more connected to instrumental conditions can be removed, still obtaining good results.

\begin{table}[H]
  \begin{center}
    \caption{Performance metrics for AGN-galaxy classification with
    directly source-related
    features only.  Spectroscopic redshifts were included in top table, photometric redshift in the middle table, no redshifts were included in bottom table). SM indicates that the training sample was built with oversampling of the minority class (see text). Obs indicates that the training sample only includes observed data. $P_{AGN}$= precision calculated with respect to the AGN class.	 $P_{g}$= precision for galaxies class. $R_{AGN}$ = recall of AGN.  $R_{g}$ = recall of galaxies. $F1_{AGN}$ = F1 score of AGN. $F1_g$ = F1 score of galaxies. ba = balanced accuracy.
     RF, Tree, AB, grad, and vote are defined in Section \ref{algorithms}.}
    \label{tab:resAGNgalphys}
   Spectroscopic redshifts\\
\resizebox{.48\textwidth}{!}{    
    \begin{tabular}{l|c|c|c|c|c|c|c}
          \hline
algorithm	&	 $P_{AGN}$ 	&	 $P_{g}$ 	&	  $R_{AGN}$	&	 $R_{g}$ 	&	 $F1_{AGN}$ 	&	 $F1_g$   & ba\\
SM & & & & & & &  \\
RF	&	0.948	&	0.739	&	0.917	&	0.823	&	0.933	&	0.779	&	0.87 \\
Tree	&	0.95	&	0.725	&	0.910	&	0.830	&	0.93	&	0.774	&	0.87 \\
\rowcolor{LightCyan}
AB	&	0.952	&	0.739	&	0.916	&	0.838	&	0.934	&	0.785	&	0.877 \\
grad	&	0.953	&	0.738	&	0.915	&	0.841	&	0.934	&	0.786	&	0.878 \\
vote	&	0.953	&	0.741	&	0.916	&	0.841	&	0.934	&	0.788	&	0.878 \\
\hline
obs & & & & & & &  \\
RF	&	0.933	&	0.788	&	0.942	&	0.761	&	0.937	&	0.774	&	0.852 \\
Tree	&	0.933	&	0.769	&	0.935	&	0.765	&	0.934	&	0.766	&	0.85 \\
AB	&	0.932	&	0.791	&	0.943	&	0.758	&	0.938	&	0.774	&	0.851 \\
grad	&	0.933	&	0.788	&	0.942	&	0.762	&	0.938	&	0.775	&	0.852 \\
vote	&	0.933	&	0.792	&	0.944	&	0.765	&	0.938	&	0.779	&	0.854 \\
\hline
    \end{tabular}
}\\
 Photometric redshifts
\\
\resizebox{.48\textwidth}{!}{
    \begin{tabular}{l|c|c|c|c|c|c|c}
      \hline
algorithm	&	 $P_{AGN}$ 	&	 $P_{g}$ 	&	  $R_{AGN}$	&	 $R_{g}$ 	&	 $F1_{AGN}$ 	&	 $F1_g$   & ba\\
SM & & & & & & &  \\
RF	&	0.96	&	0.834	&	0.937	&	0.893	&	0.948	&	0.860	&	0.915 \\
Tree	&	0.959	&	0.797	&	0.918	&	0.887	&	0.938	&	0.837	&	0.902 \\
AB	&	0.956	&	0.811	&	0.927	&	0.885	&	0.941	&	0.844	&	0.906 \\
grad	&	0.96	&	0.814	&	0.927	&	0.903	&	0.943	&	0.853	&	0.915 \\
vote	&	0.960	&	0.822	&	0.93	&	0.920	&	0.945	&	0.870	&	0.920 \\
\hline
obs & & & & & & &  \\
RF	&	0.952	&	0.838	&	0.941	&	0.869	&	0.946	&	0.851	&	0.905 \\
Tree	&	0.966	&	0.796	&	0.916	&	0.910	&	0.940	&	0.846	&	0.913 \\
AB	&	0.948	&	0.814	&	0.933	&	0.854	&	0.940	&	0.832	&	0.893 \\
grad	&	0.949	&	0.825	&	0.936	&	0.866	&	0.942	&	0.842	&	0.901 \\
vote	&	0.962	&	0.830	&	0.937	&	0.890	&	0.949	&	0.862	&	0.915 \\
\hline
    \end{tabular}
}\\
 No redshifts
\\
\resizebox{.48\textwidth}{!}{
    \begin{tabular}{l|c|c|c|c|c|c|c}
      \hline
algorithm	&	 $P_{AGN}$ 	&	 $P_{g}$ 	&	  $R_{AGN}$	&	 $R_{g}$ 	&	 $F1_{AGN}$ 	&	 $F1_g$   & ba\\
SM & & & & & & &  \\
RF	&	0.841	&	0.510	&	0.888	&	0.408	&	0.864	&	0.453	&	0.648 \\
Tree	&	0.842	&	0.540	&	0.903	&	0.400	&	0.871	&	0.459	&	0.652 \\
AB	&	0.845	&	0.587	&	0.919	&	0.405	&	0.880	&	0.479	&	0.662 \\
grad	&	0.846	&	0.583	&	0.916	&	0.413	&	0.880	&	0.483	&	0.664 \\
vote	&	0.841	&	0.590	&	0.920	&	0.400	&	0.881	&	0.468	&	0.657 \\
\hline
obs & & & & & & &  \\
RF	&	0.831	&	0.604	&	0.939	&	0.329	&	0.882	&	0.426	&	0.634 \\
Tree	&	0.827	&	0.669	&	0.959	&	0.294	&	0.888	&	0.408	&	0.626 \\
AB	&	0.822	&	0.774	&	0.979	&	0.250	&	0.893	&	0.378	&	0.615 \\
grad	&	0.823	&	0.758	&	0.977	&	0.258	&	0.893	&	0.384	&	0.617 \\
vote	&	0.826	&	0.734	&	0.971	&	0.278	&	0.893	&	0.403	&	0.624 \\
    \end{tabular}
}\\
  \end{center}
\end{table}

For test case 2, we made the same experiment to further investigate the importance of features used to train the ML algorithms.
 The results are still very similar to the ones shown for the full set of features (comparing Table \ref{tab:restype1type2phys} with Table \ref{tab:restype1type2}). This proves that the model is able to generalise beyond the specific instrumental conditions of our training dataset.
\begin{table}[H]
  \begin{center}
    \caption{Performance metrics for type 1/2  classifier with
    directly source-related
    features. Spectroscopic redshifts are included in the top table, photometric redshifts in the middle table, no redshifts are included in the bottom table. SM indicates that the training sample was built with oversampling of the minority class (see text). Obs indicates that the training sample only includes observed data.
$P_{type1}$= precision calculated with respect to the type 1 AGN class.	 $P_{type2}$= precision for type 2 AGN class. $R_{type1}$ = recall of type 1 AGN class.  $R_{type2}$ = recall of type 2 AGN class. $F1_{type1}$ = F1 score of type 1. $F1_{type2}$ = F1 score of type 2. 
    ba = balanced accuracy.
 RF, Tree, AB, grad, and vote are defined in Section \ref{algorithms}.}
    \label{tab:restype1type2phys}
    Spectroscopic redshifts\\
\resizebox{.48\textwidth}{!}{    
    \begin{tabular}{l|c|c|c|c|c|c|c}
          \hline
algorithm 	&	 $P_{type1}$ 	&	 $P_{type2}$ 	&	  $R_{type1}$	&	 $R_{type2}$ 	&	 $F1_{type1}$ 	&	 $F1_{type2}$   & ba\\
SM & & & & & & &  \\
RF	&	0.955	&	0.706	&	0.928	&	0.799	&	0.941	&	0.749	&	0.864 \\
Tree	&	0.959	&	0.681	&	0.917	&	0.816	&	0.937	&	0.742	&	0.867 \\
\rowcolor{LightCyan}
AB	&	0.962	&	0.688	&	0.918	&	0.832	&	0.94	&	0.753	&	0.875 \\
grad	&	0.963	&	0.685	&	0.917	&	0.836	&	0.939	&	0.753	&	0.876 \\
vote	&	0.961	&	0.698	&	0.922	&	0.830	&	0.940	&	0.754	&	0.873 \\
\hline
obs & & & & & & &  \\
RF	&	0.938	&	0.77	&	0.954	&	0.71	&	0.946	&	0.738	&	0.832 \\
Tree	&	0.94	&	0.762	&	0.952	&	0.717	&	0.946	&	0.739	&	0.835 \\
AB	&	0.939	&	0.779	&	0.956	&	0.712	&	0.947	&	0.744	&	0.834 \\
grad	&	0.939	&	0.78	&	0.957	&	0.71	&	0.947	&	0.743	&	0.834 \\
vote	&	0.941	&	0.782	&	0.956	&	0.723	&	0.949	&	0.751	&	0.84 \\
\hline
    \end{tabular}
}\\
 Photometric redshifts
\\
\resizebox{.48\textwidth}{!}{
    \begin{tabular}{l|c|c|c|c|c|c|c}
      \hline
algorithm 	&	 $P_{type1}$ 	&	 $P_{type2}$ 	&	  $R_{type1}$	&	 $R_{type2}$ 	&	 $F1_{type1}$ 	&	 $F1_{type2}$   & ba\\
SM & & & & & & &  \\
RF	&	0.971	&	0.804	&	0.952	&	0.864	&	0.961	&	0.826	&	0.908 \\
Tree	&	0.965	&	0.767	&	0.944	&	0.845	&	0.954	&	0.796	&	0.895 \\
AB	&	0.965	&	0.771	&	0.942	&	0.850	&	0.953	&	0.797	&	0.896 \\
grad	&	0.969	&	0.770	&	0.944	&	0.866	&	0.956	&	0.809	&	0.905 \\
vote	&	0.970	&	0.772	&	0.944	&	0.870	&	0.955	&	0.813	&	0.907 \\
\hline
obs & & & & & & &  \\
RF	&	0.967	&	0.825	&	0.961	&	0.844	&	0.964	&	0.83	&	0.902 \\
Tree	&	0.962	&	0.796	&	0.953	&	0.838	&	0.957	&	0.805	&	0.896 \\
AB	&	0.95	&	0.782	&	0.954	&	0.772	&	0.951	&	0.767	&	0.863 \\
grad	&	0.958	&	0.785	&	0.952	&	0.812	&	0.955	&	0.791	&	0.882 \\
vote	&	0.964	&	0.805	&	0.955	&	0.840	&	0.959	&	0.817	&	0.898 \\
\hline
    \end{tabular}
}
 No redshifts
\\
\resizebox{.48\textwidth}{!}{
    \begin{tabular}{l|c|c|c|c|c|c|c}
      \hline
algorithm	&	 $P_{type1}$ 	&	 $P_{type2}$ 	&	  $R_{type1}$	&	 $R_{type2}$ 	&	 $F1_{type1}$ 	&	 $F1_{type2}$   & ba\\
SM & & & & & & &  \\
RF	&	0.861	&	0.426	&	0.905	&	0.326	&	0.882	&	0.369	&	0.615 \\
Tree	&	0.858	&	0.54	&	0.949	&	0.275	&	0.901	&	0.364	&	0.612 \\
AB	&	0.859	&	0.623	&	0.965	&	0.265	&	0.909	&	0.372	&	0.615 \\
grad	&	0.859	&	0.606	&	0.962	&	0.269	&	0.907	&	0.372	&	0.615 \\
vote	&	0.858	&	0.580	&	0.955	&	0.265	&	0.907	&	0.367	&	0.613 \\
\hline
obs & & & & & & &  \\
RF	&	0.854	&	0.648	&	0.972	&	0.235	&	0.910	&	0.344	&	0.603 \\
Tree	&	0.851	&	0.762	&	0.986	&	0.203	&	0.914	&	0.320	&	0.595 \\
AB	&	0.849	&	0.861	&	0.993	&	0.184	&	0.916	&	0.303	&	0.589 \\
grad	&	0.849	&	0.859	&	0.993	&	0.184	&	0.916	&	0.303	&	0.589 \\
vote	&	0.851	&	0.830	&	0.992	&	0.199	&	0.915	&	0.323	&	0.596 \\

    \end{tabular}
    }
  \end{center}
\end{table}

\bsp	
\label{lastpage}
\end{document}